\documentclass[12pt,3p,preprint,sort&compress]{elsarticle}
\pdfoutput=1
\usepackage[colorlinks=true]{hyperref}

\usepackage{amsmath}
\usepackage{amssymb}
\usepackage{mathtools}

\renewcommand{\d}{\partial}
\renewcommand{\l}{\left(}
\renewcommand{\r}{\right)}

\newcommand{\be}{\begin{equation}}
\newcommand{\ee}{\end{equation}}
\newcommand{\ba}{\begin{align}}
\newcommand{\ea}{\end{align}}
\newcommand{\bg}{\begin{gather}}
\newcommand{\eg}{\end{gather}}
\newcommand{\bseq}{\begin{subequations}}
\newcommand{\eseq}{\end{subequations}}

\newcommand{\half}{\frac{1}{2}}

\begin{document}
\begin{flushright}
	INR-TH-2020-014
\end{flushright}

\title{Heavy Neutral Leptons from kaon decays in the SHiP experiment} 
\author[inr,mpti]{Dmitry Gorbunov}
\ead{gorby@ms2.inr.ac.ru}
\author[inr]{Igor Krasnov}
\ead{iv.krasnov@physics.msu.ru}
\author[inr,mpti,mephi]{Yury Kudenko}
\ead{kudenko@inr.ru}
\author[inr,cea]{Sergey Suvorov}
\ead{suvorov@inr.ru}
\address[inr]{Institute for Nuclear Research of Russian Academy of Sciences, 117312 Moscow, Russia}
\address[mpti]{Moscow Institute of Physics and Technology, 141700 Dolgoprudny, Russia}
\address[mephi]{National Research Nuclear University (MEPhI), 115409 Moscow, Russia}
\address[cea]{IRFU, CEA Saclay, 91191 Gif-sur-Yvette, France}
\begin{abstract}
We calculate the signal rate of hypothetical heavy neutral leptons (HNL or sterile neutrinos) from kaon decays expected in the framework of the SHiP experiment. The kaons are produced in the hadronic shower initiated in the beam-dump mode by 400\,GeV protons from CERN SPS. For a sufficiently light HNL (when the decays are kinematically allowed) we find kaon decays to be a noticeably richer source of HNL as compared to $D$-meson decays adopted in previous studies of the HNL phenomenology at SHiP. In particular, SHiP is capable of fully exploring the central part of the kinematically allowed region of the HNL mass and mixing with  electron and muon neutrinos down to the lower cosmological bound. The latter is associated with HNL decays in the early Universe to  energetic products rescattering off and thus destroying light nuclei produced at the primordial nucleosynthesis. A consistency of the HNL model with smaller mixing would require either a hierarchy -- much larger mixing of all the HNL with tau neutrino -- or 
non-standard cosmology and new ingredients in the HNL sector, closing the room for the minimal non-seesaw type I model with sterile neutrinos lighter than kaons.         
\end{abstract}

\date{}

\maketitle

{\bf 1.} The experiment SHiP (Search for Hidden  Particles) \cite{Anelli:2015pba,SHiP:2018yqc} on the base of the CERN SPS proton beam has been proposed\,\cite{Bonivento:2013jag} with primary goal of searching for hypothetical feebly interacting light particles. Many well-motivated examples can be found in Ref.\,\cite{Alekhin:2015byh} highlighting the SHiP physics case. SPS protons of 400\,GeV hit the Molybdenum-Tungsten target, producing secondary particles, which subsequently rescatter and decay forming hadronic and electromagnetic showers degrading downstream in a target. The interesting hypothetical particles can emerge from the target and, being feebly interacting, propagate freely. On the contrary, the host of ordinary, Standard Model (SM) particles must be either absorbed in the downstream dumper or deflected (mostly muons) by  specially designed magnets. The SHiP target complex is optimized to obtain  a flux of hypothetical particles (if any) with as little contamination by the ordinary particles as possible. The hypothetical particles then may show up downstream in the detector system by decaying into SM particles, which make the signal events. 

While this scheme is fairly universal giving a good chance to explore many extensions of the SM with feebly interacting light particles, it has been chosen in \cite{Bonivento:2013jag} based on the seemingly best performance in direct searches for {\it sterile neutrinos} --- massive fermions, singlet with respect to the SM gauge group, which mix with active neutrinos --- dubbed heavy neutral leptons (HNL). These particles, if sufficiently light, can be produced via mixing mostly in decays of mesons and similarly decay into lighter SM particles. If the mixing is small, a beam-dump facility is the best place to search for these particles\,\cite{Gorbunov:2007ak}. The heavier the meson, the bigger the room for HNL mass to be investigated. However, presently available in a beam-dump mode proton beams are not sufficiently energetic to produce beauty mesons in the required amount, and most promising candidates to be the HNL sources in the beam-dump experiment are $D$-mesons. Based on this idea a new dedicated experiment at CERN has been proposed\,\cite{Gninenko:2013tk}, and its structure, detector geometry and position were chosen\,\cite{Bonivento:2013jag} to get the highest signal rate of HNL from $D$-mesons produced in a fixed-target experiment by protons  at  CERN SPS. The estimate of this signal was further revised in \cite{Alekhin:2015byh} and later in  \cite{SHiP:2018xqw}, motivated by both the project natural development and theoretical progress\,\cite{Bondarenko:2018ptm}. 

While $D$-mesons can produce HNLs with masses up to about 2\,GeV, the HNL of smaller masses, up to about 0.5\,GeV and 0.14\,GeV can be  also produced  in decays of kaons and pions,  respectively. So far these production modes have been fully neglected. The reason is natural: in the target material the kaon and pion decay lengths grossly exceed their interaction lengths. Consequently, these light mesons get absorbed and certainly degrade in energy. The 3-momenta of produced HNL are generically not pointed towards the detector, and their typical values are too small to pass the selection criteria for signal  events and  to suppress the background. Nevertheless, this question has not been studied before, and here we present our estimates for the signal rates of {\it HNL appeared in kaon decays} inside the SHiP target. We observe, that  disadvantages of the mildly-relativistic kinematics are compensated to large extent by a high kaon production rate. We find this result encouraging, since the expected signal rate at the SHiP detector is not only noticeably higher than that for HNL (of the same mass) from $D$-mesons, but also high enough to fully explore a large part of the  allowed by the Big Bang Nucleosynthesis region  (in models without the special hierarchy: mixing with the tau neutrino should not strongly dominate). Inside it  HNL decays in the early Universe do not spoil the production of light nuclei in the cosmic plasma (energetic SM particles from HNL decays might be dangerous, destroying the light nuclei). This finding strengthens the physics case of SHiP, which results may be of great importance for the seesaw type I mechanism  \cite{Minkowski:1977sc,Yanagida:1980xy} of generating active neutrino masses.  

The paper is organized as follows. We start with the model description explaining possible origin of the  HNL mixing with active neutrinos and introduce the required notations. Then we outline the part of SHiP facility relevant for our study: the target structure, the position and geometry of the main detector. Further we turn to the kaon production by the beam protons on target, that we calculate by making use of the GEANT4 package\,\cite{Agostinelli:2002hh}. Then we trace kaon decays into HNL and the latter subsequent decays inside the detector volume to  the SM particles forming the signal events. Applying the cuts, presently accepted by the SHiP collaboration, as well as the zero-background hypothesis, we calculate the number of signal events and the corresponding limits on the sterile-active neutrino mixing to be placed by the SHiP experiment in case of no evidence. Finally, we summarize with discussion of possible options to improve the SHiP performance in searches for HNL.  

{\bf 2.} In this study we follow a pure phenomenological approach, which guarantees a universal result, relevant for various models with HNL. Namely, to the Standard Model (SM) we introduce one hypothetical fermion of a mass $M_N$ smaller than the kaon mass, $M_N<M_K$. The fermion is a singlet with respect to the SM gauge group, but it mixes with active neutrinos (electron $\nu_e$, muon $\nu_\mu$, and $\tau$-neutrino $\nu_\tau$), that we parameterize with dimensionless variables $U_{\alpha}$, $\alpha=e\,,\mu\,,\tau$. Considering the models with $|U_{\alpha}|\ll 1$ and $M_N$ much exceeding the scale of  the active neutrino mass, one can treat this fermion as the fourth massive neutrino (that explains the notation in the mixing parameter and the very term {\it HNL}), but not as the neutrino of a hypothetical fourth SM generation (that explains the alternative name {\it sterile neutrino}). 

From the phenomenological side we indeed have in the model a heavy neutrino, which participates in all  weak processes (if kinematically allowed) but with effective coupling  of the weak gauge constant multiplied by the corresponding parameter $U_{\alpha }$. This determines both the HNL production and its decay. In the absence of any other interactions this fermion is an example of a feebly interacting particles.  

From the theoretical side, this construction can be found in many extensions of the SM. Apparently, the simplest yet self-contained example is provided by the sterile neutrino model in the framework of the see-saw type I mechanism \cite{Minkowski:1977sc,Yanagida:1980xy}. There the NHLs $N_I$, $I=1,2,\dots$ are Majorana fermions with the Lagrangian 
\[
{\cal L}=i\bar N_I\gamma^\mu\d_\mu N_I-\l Y_{\alpha I}\bar L_\alpha \tilde H N_I-\half M_I\bar N_I^c N_I+\text{h.c.}\r. 
\]
Here $Y_{\alpha I}$ and $M_I$ are dimensionless Yukawa couplings and Majorana masses, $L_\alpha$ are SM lepton doublets, and $\tilde H=\epsilon\times H^*$ is the conjugated SM Higgs doublet (the weak indices are omitted). The Higgs field gains  a non-zero vacuum expectation value, which yields a mixing mass term for $N_I$ and active neutrinos $\nu_\alpha$. Upon diagonalization the active neutrinos acquire masses (that explains neutrino oscillations), and HNL mass states mix with active neutrinos. If all $M_I\gg 1$\,eV, the mixing is small and original $N_I$ are almost identical to the HNL mass states. At the same condition, the active neutrino mass scale is suppressed by mixing squared (the see-saw mechanism at work), which may explain why this scale is much lower than any other mass of the SM particles.   

At least three sterile neutrinos are needed to provide masses to all three active neutrino states. The sterile neutrino masses are free parameters in between 1\,eV and $10^{14}$\,GeV (the latter scale refers to the perturbativity limit of $|Y_{\alpha I}|<1$), for details see e.g.\,\cite{Gorbunov:2014efa}. If some of the sterile states are lighter than the kaon (but significantly heavy to decay into  charged SM particles), the corresponding phenomenology at SHiP is exactly the same as we consider in this paper. A particular model can predict not one, but two, three, etc HNL states in the interesting mass range. Then the mixing parameters $U_{\alpha I}$ can be appropriately constrained and related. We do not study this case, but our results can be straightforwardly modified to cover it. In a similar way, one can refine our limits for the case with degenerate sterile neutrinos, where the detector fails to distinguish two (three, etc) close in mass states. This situation is realized in $\nu$MSM (for a review see \cite{Boyarsky:2009ix}), where two sterile neutrinos are highly degenerated to explain both phenomena:  neutrino oscillations and the  baryon asymmetry of the Universe. The mixing pattern, $|U_{\alpha I}|$, and the corresponding signals are specifically constrained in this case\,\cite{Gorbunov:2007ak}. 

In what follows we consider the model with one HNL of the mass $M_N\ll M_K$ and small and unrelated mixing parameters $|U_{\alpha}|\ll 1$, which  are additionally bounded  from above for  realistic models by direct searches for HNL, for a recent summary see e.g.\,\cite{Chun:2019nwi}. 

{\bf 3.} In the SHiP experiment a 400\,GeV proton beam hits a target, where secondary particles form hadronic showers dominated by pions and kaons. When a kaon decays it has a chance (proportional to the corresponding differential branching ratio) to produce a sterile neutrino of a given 3-momentum. For  a heavy neutrino two-body kaon decays are not suppressed by the chirality flip, so they dominate over  three-body decays because of the phase volume, see formulas and plots in Ref.\,\cite{Gorbunov:2007ak}. If HNL is heavier than a pion the domination factor is at least 50 and typically much higher.  Moreover, the three-body decay is relevant only for more narrow range of HNL masses, $M_N<M_K-M_\pi$, due to kinematics. Therefore, we account only for charged kaons and consider only their two-body decays as the HNL sources. 

The HNL lives long and interacts feebly, hence it propagates freely along its 3-momentum at production. To be observed at SHiP, its trajectory must pass through the decay volume, which is placed at a distance of $L=64$\,m from the target. The decay volume is of $\Delta l=50$\,m length and has a conical form along the beam with a rectangular base\,\cite{SHiP:2018xqw}: the near and far ends have heights of $\Delta h_1=4.5$\,m and $\Delta h_2=10$\,m and widths of $\Delta w_1=2.4$\,m and $\Delta w_2=5$\,m, respectively. Parameters of the SHIP decay volume and number of protons on target  (POT) used in our study  are shown in Tab.\,\ref{tab:initial}.
\begin{table}[!htb]
\begin{center}
\begin{tabular}{|l |l |}
\hline
Protons on target & \,$N_{POT} = 2 \times 10^{20}$\\
\hline
Decay volume length & \,$\Delta l =$ 50 m\\
\hline
Decay volume upstream height & \,$\Delta h_1 =$ 4.5 m\\
\hline
Decay volume upstream width & \,$\Delta w_1 =$ 2.4 m\\
\hline
Decay volume downstream height & \,$\Delta h_2 =$ 10 m\\
\hline
Decay volume downstream width & \,$\Delta w_2 =$ 5 m\\
\hline
Decay volume distance from target & \,$L =$ 64 m\\
\hline
\end{tabular}
\caption{Parameters of the SHIP decay volume.}
\label{tab:initial}
\end{center}
\end{table}
The fiducial volume of the SHiP detector almost coincides with its decay volume. 

To be observed, the HNL must decay inside the detector volume. The background-free signatures assume (at least) two charged particles in the final state, each with  3-momenta of  $\geq$1\,GeV/c. Both particles must cross the far end of the decay  volume to be registered by the SHIP decay spectrometer. The detection efficiency  is expected to be high, so we set it to 100\% in our study.

\begin{figure}[!htb]
  \centering
  \includegraphics[width=0.7\linewidth]{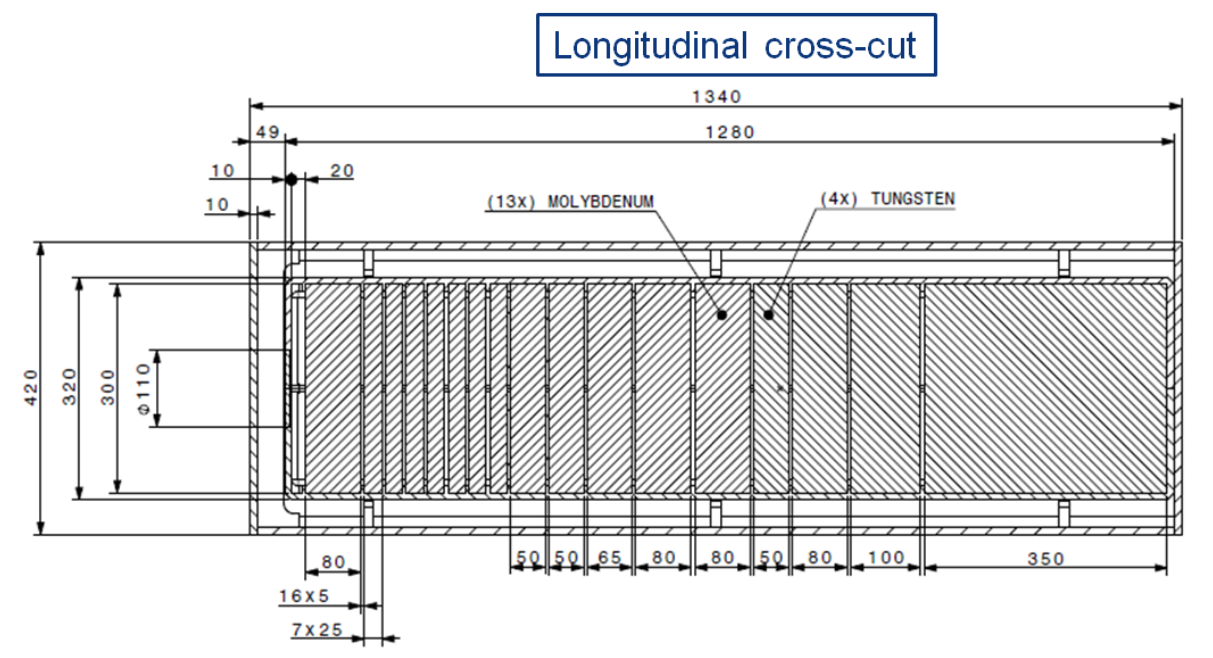}
  \caption{The schematic view of the SHiP target; the dimensions of the target are given  in mm.}
  \label{fig:target}
\end{figure}

{\bf 4.} To obtain the distribution of the HNL decay products over 3-momenta, we start with the  analysis of the kaon production and decay in the SHiP target. We simulate the kaon production by 400\,GeV protons by exploiting the GEANT4 framework\,\cite{Agostinelli:2002hh}. 
The target design is taken from \cite{Anelli:2015pba}. The schematic view  of the Molybdenum-Tungsten target with water interlayers is shown in Fig.\,\ref{fig:target}.
Light mesons emerged in proton interactions  propagate farther, mostly scattering in the target material and thus producing more mesons. The physics list we utilize for the processes in the target is called ``QGSP BERT''. It uses the quark-gluon string model for the high energy events (energy in the center-of-mass frame is $\gtrsim$ 10 GeV) and the Bertini cascade model for lower energies ($\lesssim$ 10 GeV).  All the standard EM processes and decays are accounted for.

\begin{figure}[!htb]
  \centering
    \includegraphics[width=0.7\textwidth]{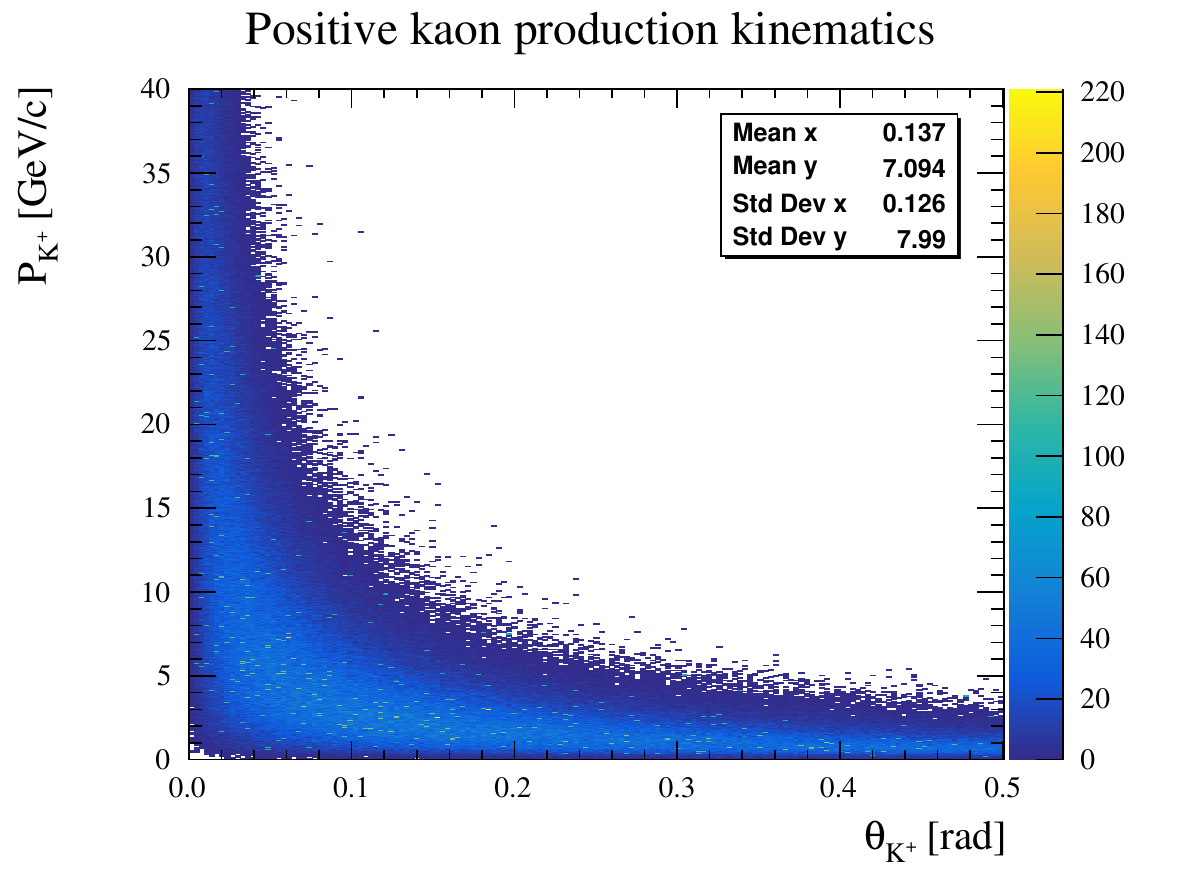}
    \caption{The spectrum of $K^+$ produced in the first collisions of 400\,GeV protons incident on the Mo-Tn target: the result of GEANT4 simulation.}
\label{fig:prod_Kpos}
\end{figure}

\begin{figure}[!htb]
  \centering
  \includegraphics[width=0.7\linewidth]{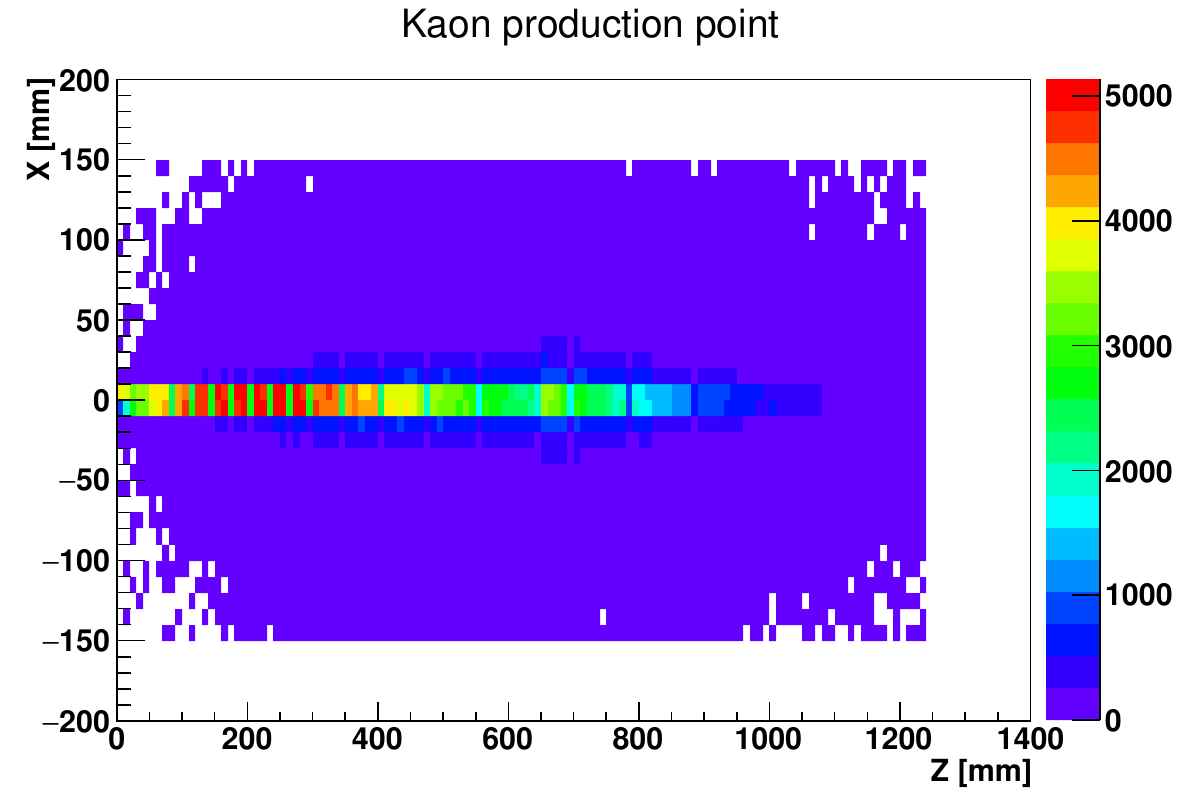}
  \caption{The distribution of charged kaon production points in the target.}
  \label{fig:kaon_prod_point}
\end{figure}

\begin{figure}[!htb]
  \centering
  \includegraphics[width=0.7\linewidth]{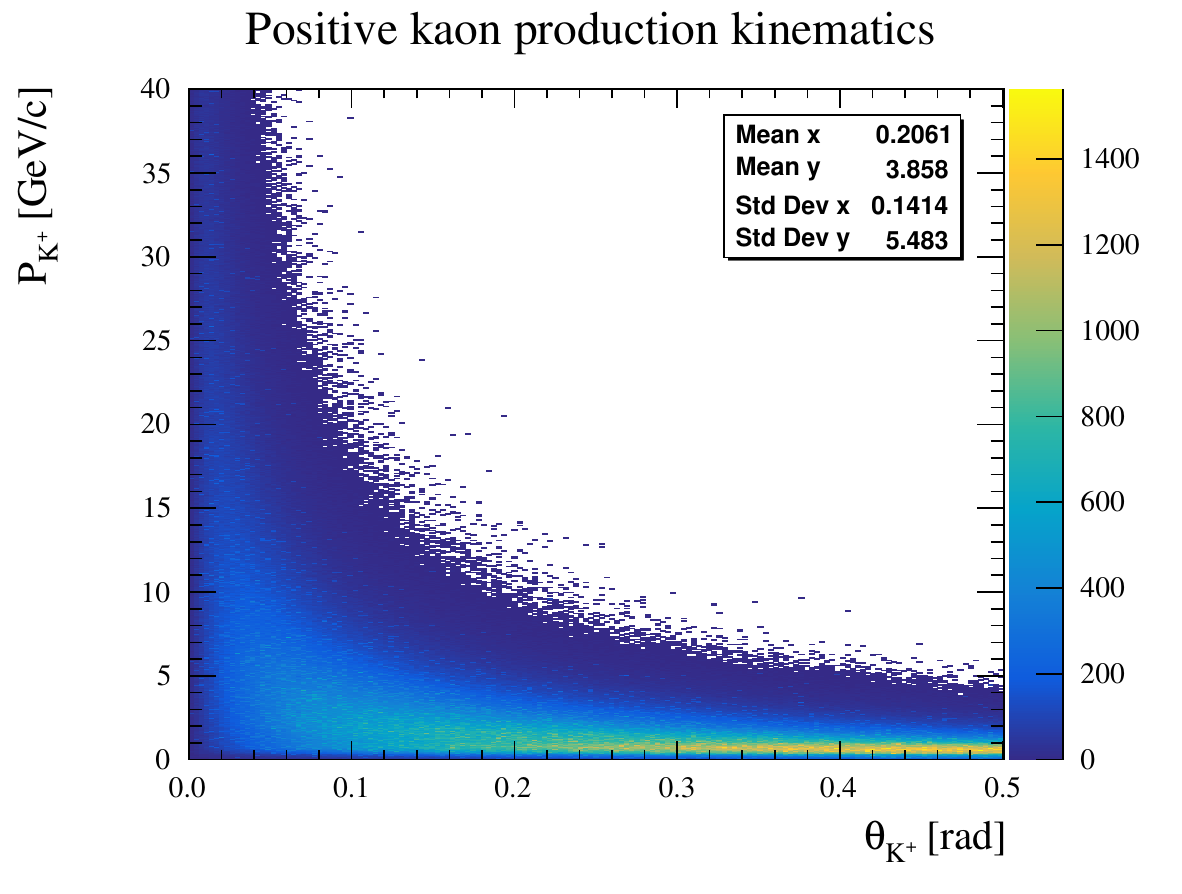}
  \caption{The overall $K^+$ spectrum, all momentum are taken at the production point; the results of the  GEANT4 simulation.}
  \label{fig:Kpos_all}
\end{figure}

Naturally, the most energetic kaons are produced in the very first collision of the  incident protons. The spectrum of these charged kaons $K^+$ is presented in 
Fig.\,\ref{fig:prod_Kpos}.
The secondary hadrons (and degraded protons) also produce kaons  inside the target. The distribution of the charged kaon production points is shown in Fig.\,\ref{fig:kaon_prod_point}. 
One observes that most of the kaons are produced in the Molybdenum layers in the upstream target region. Since secondary particles degrade in energy while the hadronic shower develops, the produced by them kaons have lower energy. The spectrum of all $K^+$ at their production points  inside the target is shown in Fig.\,\ref{fig:Kpos_all}.

After production, each koan propagates in the target, and having a strange quark in its content, can either decay via weak coupling or interact with a nuclei. In case of inelastic interactions, some charged kaons may transform to either the neutral ones or strange baryons. The former  may transform via inelastic scattering back to charged kaons or decay into non-strange particles, while the latter decay into non-strange baryons and mesons. Both neutral kaons and strange baryons can produce HNLs via semileptonic weak decays, but those are three-body decays, so the rates are significantly suppressed, and overall kinematics makes them open only for lighter HNLs. Therefore, only   charged kaons produced in the target are considered as parents of potentially interesting HNLs.

The kaon can decay at any moment, but in the laboratory frame this chance grows with kaon attenuation: most kaons decay (almost) at rest, when its 3-momentum falls below about 0.1\,MeV/c. However, in this case the decay products have no preferable directions, and only a very small part of them pass through the SHiP decay volume. Moreover, the kaon, decaying at rest, cannot produce  HNL with the total energy  $>$0.5\,GeV, that subsequently constrains the HNL decay product energy, well below the detection threshold of the   SHiP decay spectrometer. Therefore, we neglect these stopped kaons as well. 

\begin{figure}[!htb]
  \centering
  \includegraphics[width=0.7\linewidth]{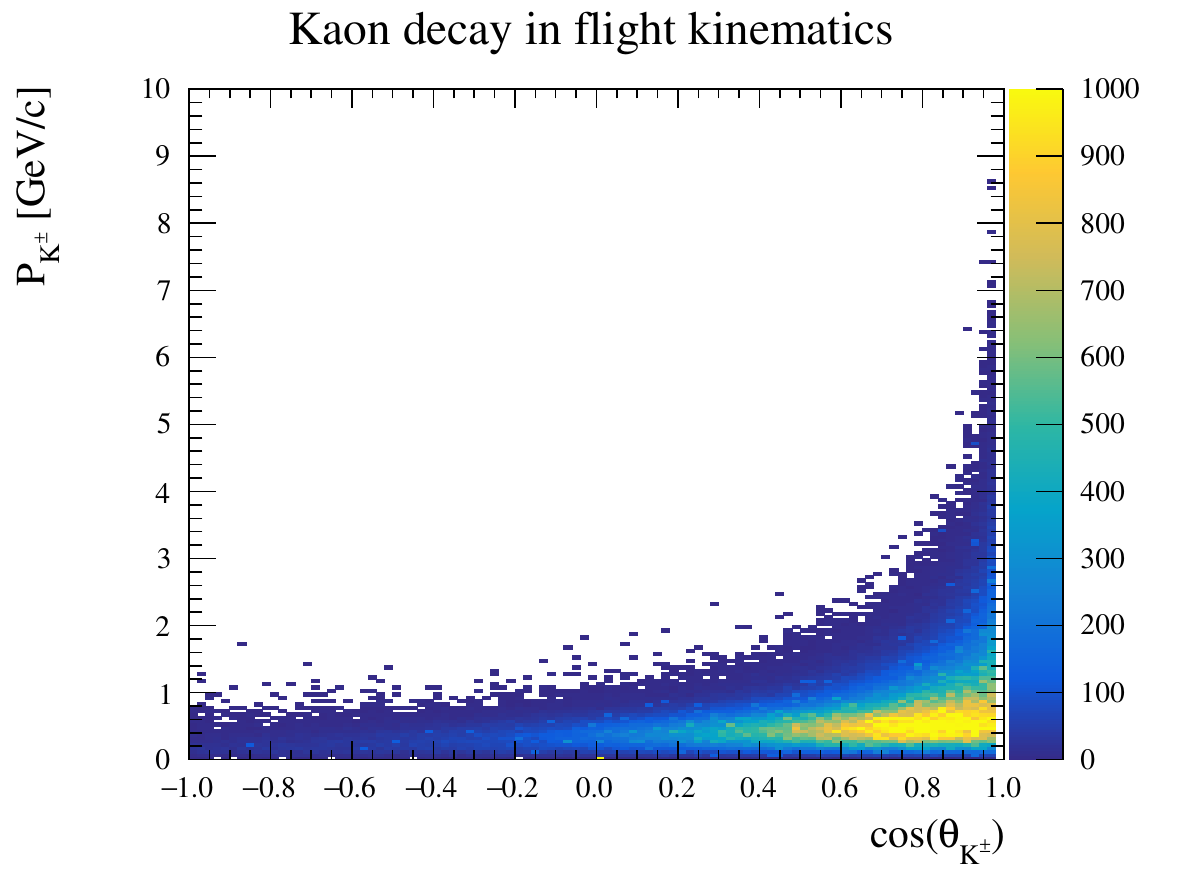}
  \caption{Kinematics of $K^\pm$ decays in flight.}
  \label{fig:kaon_spectrum}
\end{figure}

To conclude, only charged kaons decaying in flight are  relevant for HNL searches. Based on the simulated statistics of $N_{stat}=1.5\times10^6$ POT we observe that each proton produces about 8.01 $K^+$ and 3.51 $K^-$, out of which 4.52 $K^+$ and 2.85 $K^-$ are captured (while strangeness transfer mostly to neutral kaons and baryons, respectively) and 3.18 $K^+$ and 0.59 $K^-$ decay at rest. Therefore, the only profitable are 0.29 $K^+$ and 0.07 $K^-$ decaying in flight. Their     distribution over total 3-momentum and polar angle (with zenith direction along the beam line) is presented in Fig.\,\ref{fig:kaon_spectrum}.
Let us note in passing that about 0.1\% of the produced charged kaons leave the target. We neglect their contribution. Likewise, in our simulations we ignore a possible impact of the active muon shield with magnetic field (designed primary to deflect the muons outside the target) on the propagation of charged kaons inside the target. 


{\bf 5.} The described above GEANT4 simulation yields a set of coordinates $\{x_1, y_1, z_1\}$ and 3-momentum $p_K=\{p_{x_1}, p_{y_1}, p_{z_1}\}$ of the charged kaons at the moment of their decay. We take into account both $K^+$ and $K^-$. 
Since each proton in the beam produces on average $M_{p\to K}=0.36$ charged kaons that decay in flight, the  total number of in-flight kaon decays is expected  to be $K_{tot}=N_{POT}\times M_{p\to K}$. 

Each charged kaon may decay into HNL with the probability equal to the corresponding branching ratio (recall, we account only for the two-body kaon decays) \cite{Gorbunov:2007ak} 
\begin{equation}
    \label{Br-2}
    \begin{split}
&\text{Br}(K^\pm \to l^\pm N)  = |U_{l}|^2 \frac{\tau_K}{8 \pi}  G_F^2 f_K^2 M_K M_N^2 |V_{us}|^2\\
&\times\l1 - \frac{M_N^2}{M_K^2} + 2 \frac{M_l^2}{M_K^2} + \frac{M_l^2}{M_N^2} \l1-\frac{M_l^2}{M_K^2}\r\r\times \sqrt{\l 1 + \frac{M_N^2}{M_K^2} - \frac{M_l^2}{M_K^2}\r^2 - 4\, \frac{M_N^2}{M_K^2}}\,.
\end{split}
\end{equation}
Here $f_K=155.6$\,MeV and $\tau_K=1.238\times 10^{-8}$\,s are the kaon decay constant and lifetime, and $M_l$ is the charged lepton mass, $l=e,\,\mu$. The total number of HNL produced by the relevant kaons at SHiP reads 
\[
N_{N}=K_{tot}\times\sum_{l=e,\,\mu}\text{Br}(K^\pm \to l^\pm N)\,.  
\]
To use all the GEANT4 statistics, in our numerical calculations each decaying in flight charged kaon  
is assumed to produce a sterile neutrino.
In the final calculation we account for that via 
multiplying the number of ``signal events'' 
by $N_{N}/K_{tot}$. 
Unless the statistics is too low, we obtain viable results.
If our statistics proves to be insufficient, we improve it by calculating several (tens, hundreds) sterile neutrino production scenarios for each event of the kaon decay simulation.

The HNL trajectories and chances to decay inside the SHiP decay volume are obtained following the procedure of Ref.\,\cite{Krasnov:2019kdc}. The trajectory starts at the point of the kaon decay, $\{x_1, y_1, z_1\}$, provided by the simulations. In the kaon rest frame the HNL 3-momentum $p=\{p_x, p_y, p_z\}$ is randomly distributed over the dimension-2 sphere of radius fixed for a given decay mode \eqref{Br-2} as 
\begin{equation}
\label{eq:2body}
p = \frac{M_K}{2} \sqrt{\left(1 - \left(\frac{M_N+M_l}{M_K}\right)^2\right) \left(1 - \left(\frac{M_N-M_l}{M_K}\right)^2\right)}
\end{equation}

\begin{figure}[!htb]
\begin{center}
\includegraphics[width=0.7\textwidth]{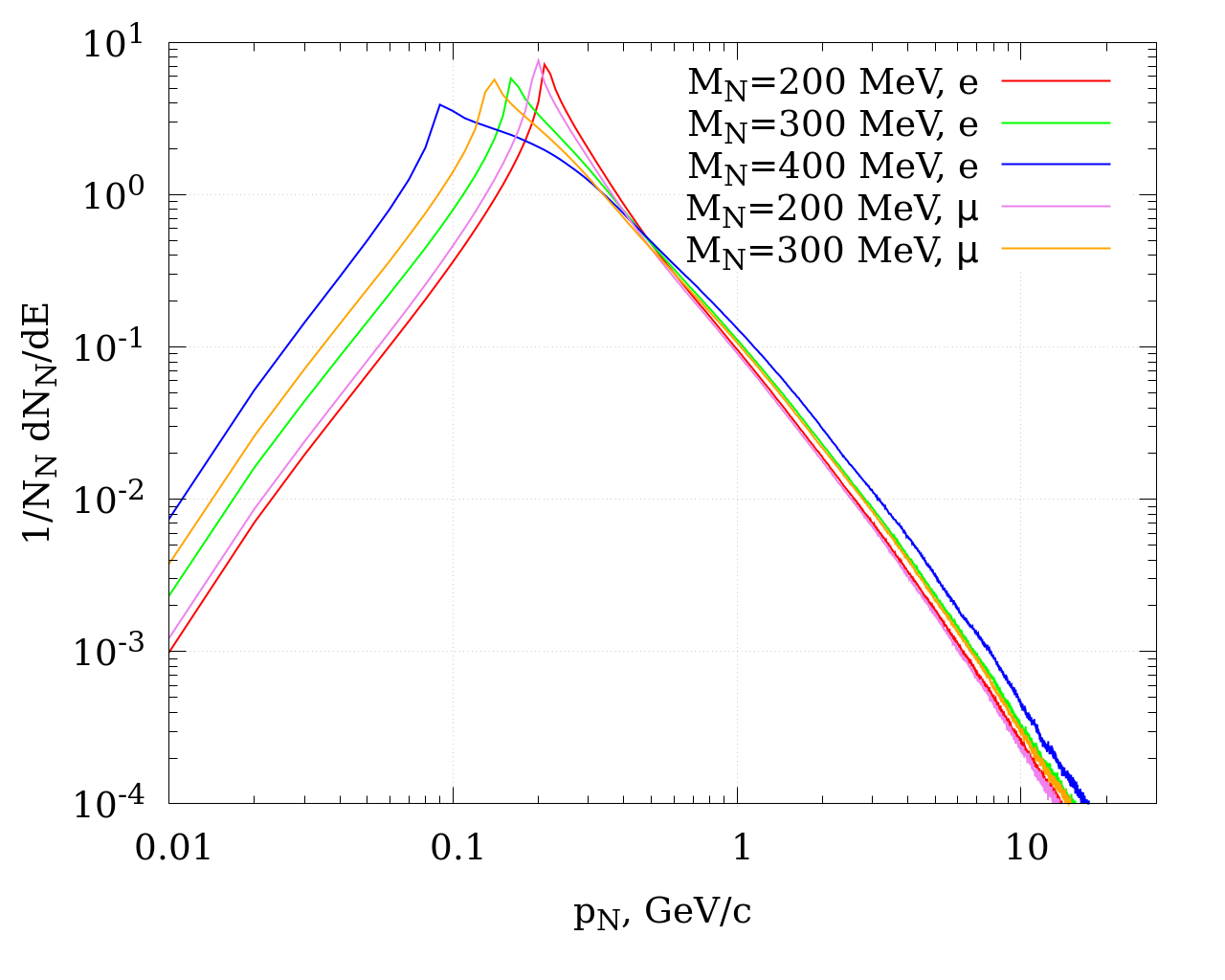}
\caption{The spectra of sterile neutrinos, produced from non-stopped charged kaons.}
\label{fig:spectrum1_non}
\end{center}
\end{figure}

To obtain the NHL 3-momentum in the laboratory frame $p_N$, we make a Lorentz boost defined by the kaon 3-momentum in the laboratory frame taken from the simulations. 
We choose the point where proton beam crosses the target as the origin of the coordinate system, $z$ axis is directed towards the detector and $x,y$ axes are chosen so that for a particular decaying kaon its longitudinal component is along $z$-axis, while the transverse component is along $x$: $p_{K_L} \equiv p_{K_z}=p_{z_1}$, $p_{K_T} \equiv p_{K_x}$, $p_{K_y}=0$. Then the components of HNL 3-momentum   
in this frame read:
\begin{eqnarray}
\label{p_Ny}
p_{N_x} & = & -\frac{E_N}{M_K} p_{K_T} - p_z \sqrt{1+\frac{p_K^2}{M_K^2}}\frac{p_{K_T}}{p_K} + p_x \frac{p_{K_L}}{p_K}\,,\\
\label{p_Nz}
p_{N_y} & = & p_y\,,\\
\label{p_Nx}
p_{N_z} & = & \frac{E_N}{M_K} p_{K_L} + p_z \sqrt{1+\frac{p_K^2}{M_K^2}} \frac{p_{K_L}}{p_K} + p_x \frac{p_{K_T}}{p_K}\,.
\end{eqnarray}
This frame is related to the laboratory frame (where the kaon 3-momentum equals $p_K=\{p_{x_1}, p_{y_1}, p_{z_1}\}$) through rotation along the $z$-axis which restores a non-zero $y$-component of the kaon momentum. Hence, to adopt formulas \eqref{p_Ny}-\eqref{p_Nx}, we perform two rotations: before the Lorentz-boost we rotate to the frame there $p_{K_y}=0$, and after the boost we rotate back to the laboratory frame. We plot the overall spectra of the obtained in this way HNL in Fig.\,\ref{fig:spectrum1_non}.
In what follows we keep for the HNL 3-momentum in the laboratory frame the same notations as in \eqref{p_Ny}-\eqref{p_Nx}, that is $p_N=\{ p_{N_x}, p_{N_y},p_{N_z}\}$.     


{\bf 6.} Next, we check, whether the HNL trajectory crosses the decay volume or not. The interesting HNL must have $p_{N_z}>0$, otherwise it flies backwards. Then, at the moment when its position is at $z_N=L$ (where the decay volume begins, see Tab.\,1), its other coordinates are
\begin{equation}
\label{eq:xN_yN}
x_N= x_1 + \frac{p_{N_x}}{p_{N_z}}( L - z_1 )\,,\;\;\;\;\;\;\;
y_N=y_1 + \frac{p_{N_y}}{p_{N_z}}( L - z_1 )\,.
\end{equation}
If HNL decays in the decay volume its coordinates are:  
\begin{eqnarray}
\label{eq:xN}
x_N&=& x_1 + \frac{p_{N_x}}{p_{N_z}}( L + \Delta l \cdot u - z_1 ),\\
\label{eq:yN}
y_N&=&y_1 + \frac{p_{N_y}}{p_{N_z}}( L + \Delta l \cdot u - z_1 ),\\
\label{eq:zN}
z_N&=&L + \Delta l \cdot u,
\end{eqnarray}
where $u \in (0,1)$ is some number. Inequalities 
\[
|x_N|< \half \l \l 1-u\r \Delta w_1+u \Delta w_2\r\,,\;\;\; 
|y_N| < \half \l \l 1-u\r \Delta h_1+u \Delta h_2\r\
\]
guarantee that the decay happens inside the decay volume. 

\begin{figure}[!htb]
\begin{center} 
\includegraphics[width=0.45\textwidth]{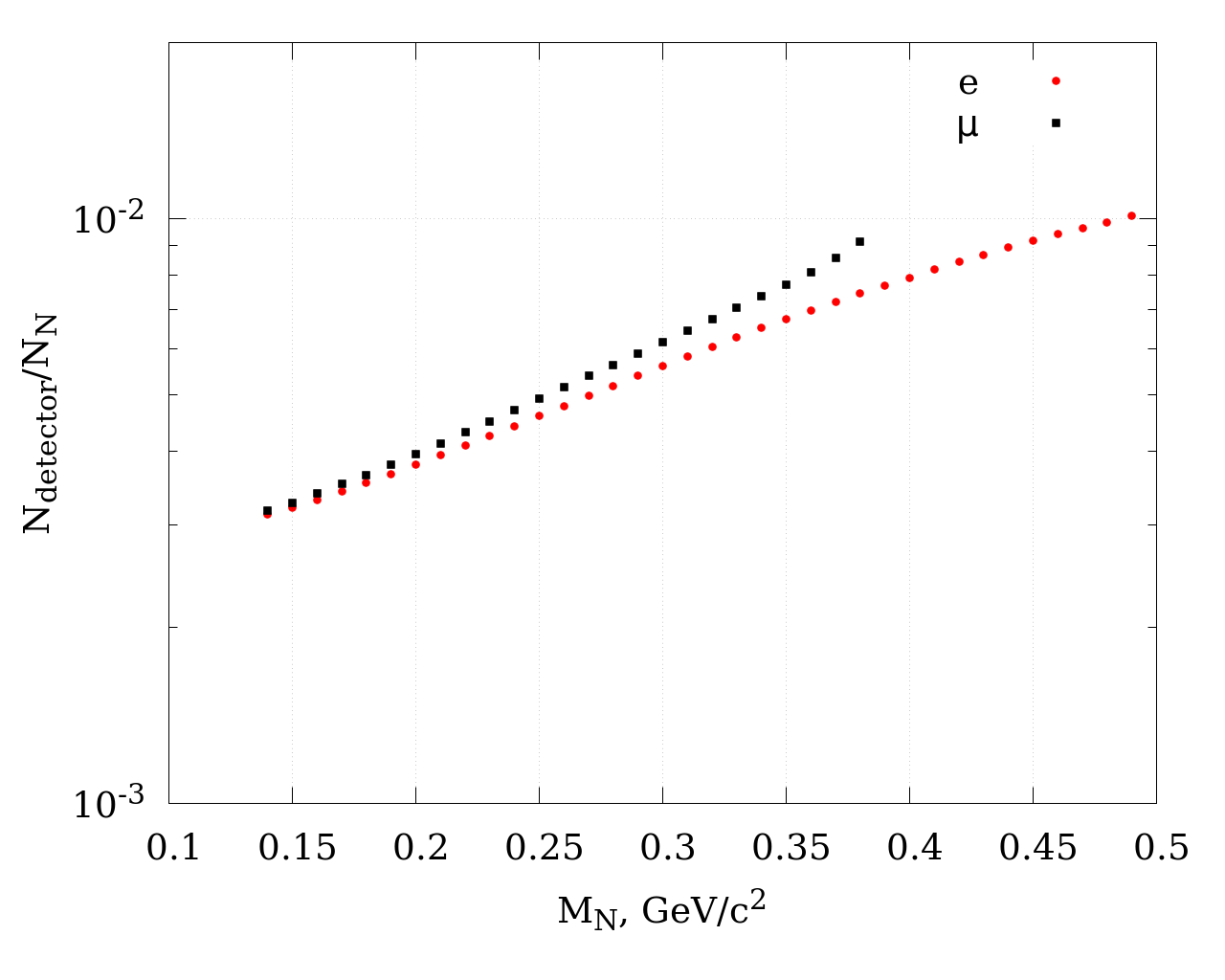}
\hskip 0.05\textwidth 
\includegraphics[width=0.45\textwidth]{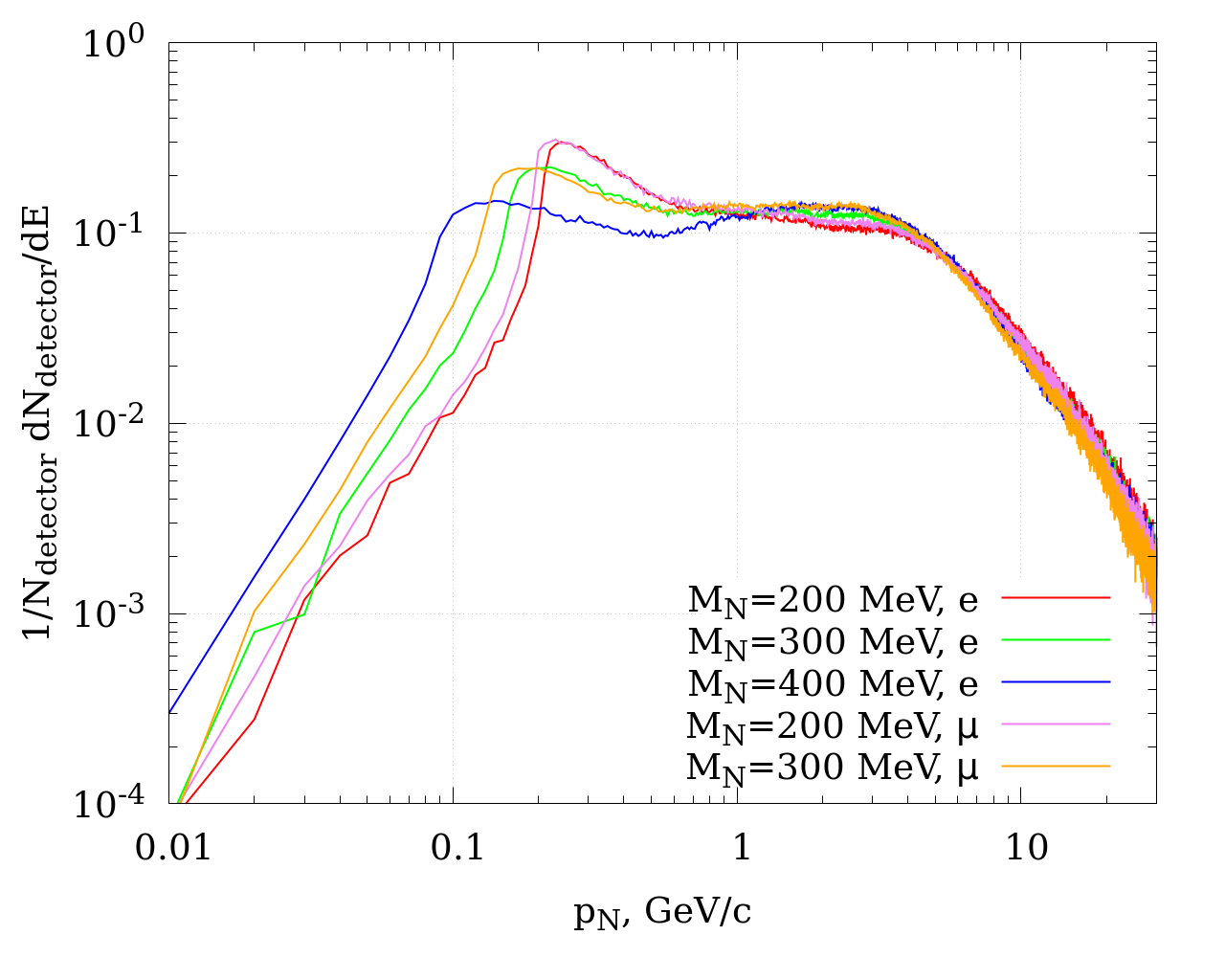}
\caption{Left panel: dependence of the fraction of the sterile neutrinos that pass through the decay volume on the sterile neutrino mass. Right panel: the spectrum of the sterile neutrinos that pass through the decay volume.}
\label{fig:N_detector_non}
\end{center}
\end{figure}

We count all neutrinos that enter the decay volume along the suitable trajectory and call the resulting number $N_{detector}$.
We present the ratios of $N_{detector}/N_N$ of sterile neutrinos reaching the decay volume on the left panel of  Fig.\,\ref{fig:N_detector_non}. 
The ratios grow with the HNL mass and finally reach about 1\%. They illustrate the signal suppression due to the geometrical factor.   The HNL spectra are shown on right panel of Fig.\,\ref{fig:N_detector_non}. 

\begin{figure}[!htb]
\begin{center}
\includegraphics[width=0.45\textwidth]{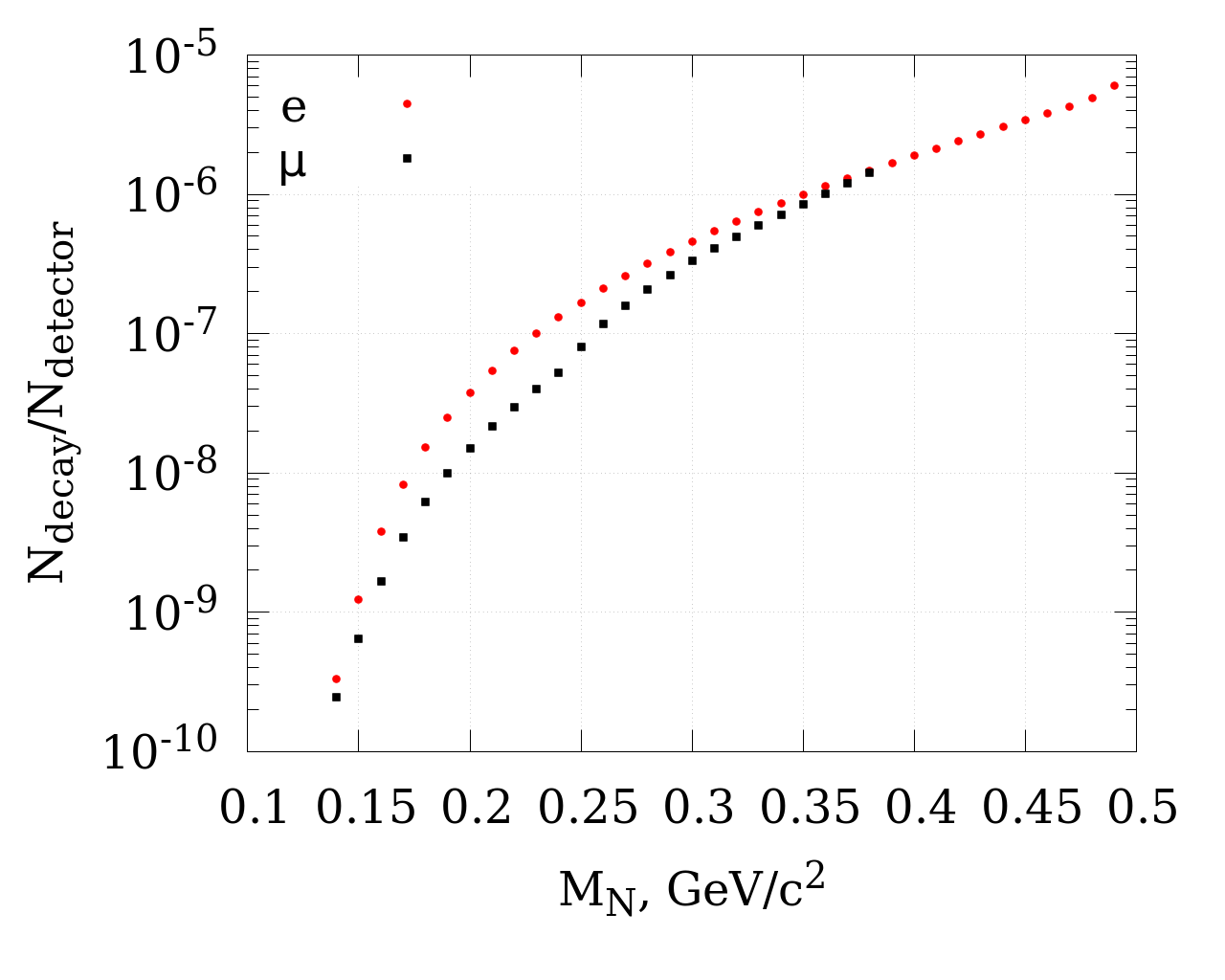}
\hskip 0.05\textwidth 
\includegraphics[width=0.45\textwidth]{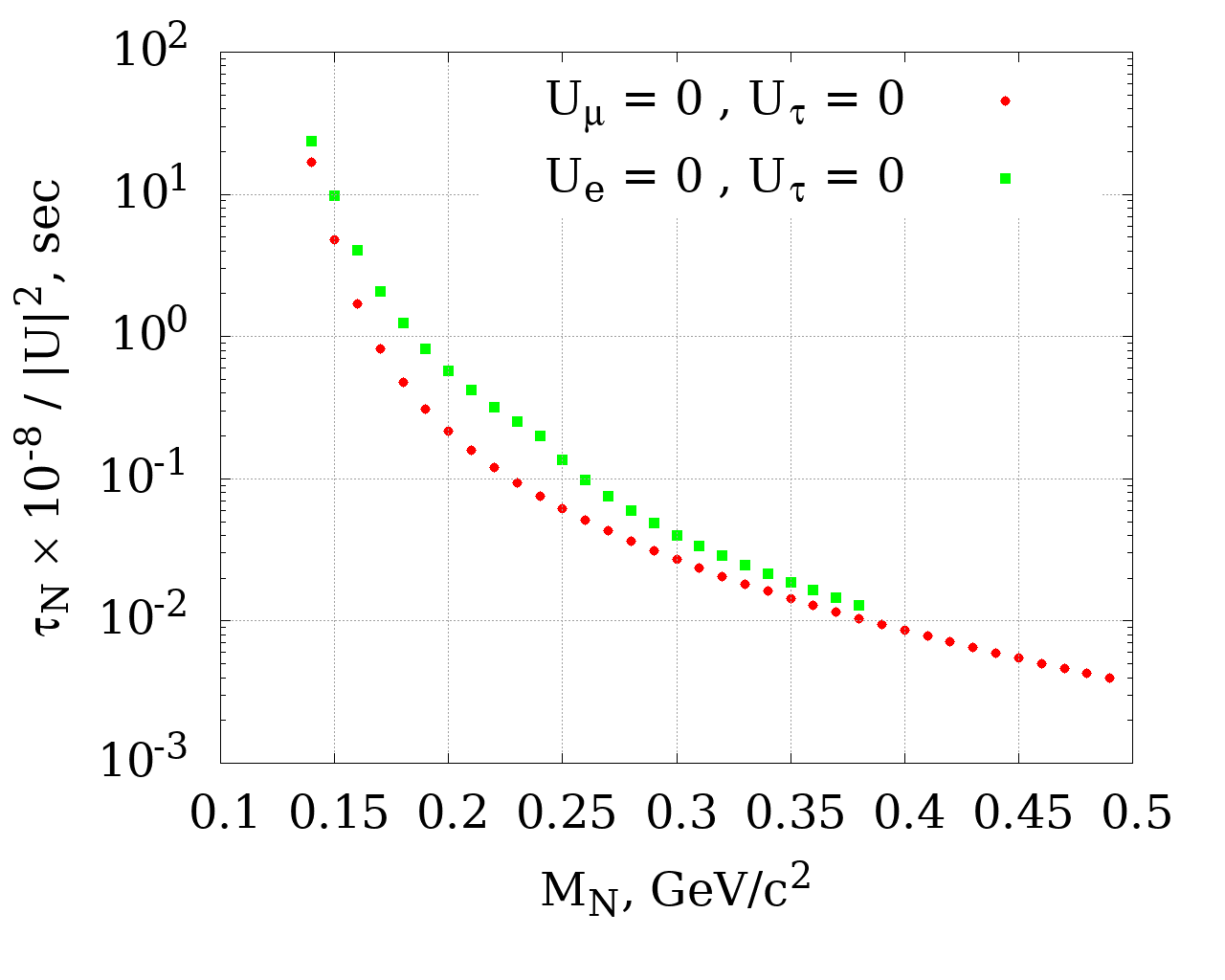}
\caption{The fraction of HNL that decay in the detector volume as a function of HNL mass (left panel). The sterile neutrino lifetime (right panel). We consider mixing with either electron or muon neutrinos, $|U|^2\equiv \sum_\alpha |U_{\alpha}|^2$. 
\label{fig:N_decay_non}}
\end{center}
\end{figure}

The probability for the sterile neutrino to decay at the moment it reaches coordinates \eqref{eq:xN} - \eqref{eq:zN} for any $u$ (that means it decays inside the decay volume) reads 
\begin{eqnarray}
\label{eq:P}
P & = & \exp{\left(-\frac{(L-z_1)}{\tau_N} \frac{M_N}{p_{N_z}}\right)} \left(1 - \int_0^1 du \exp\left(- \frac{\Delta l \cdot u}{\tau_N}\frac{M_N}{p_{N_z}}\right)\right) =\nonumber\\
&=& \exp\left(-\frac{(L-z_1)}{\tau_N} \frac{M_N}{p_{N_z}}\right) \left(1 - \frac{\left(1 - \exp\left(- \frac{\Delta l}{\tau_N}\frac{M_N}{p_{N_z}}\right)\right)}{\left( \frac{\Delta l}{\tau_N}\frac{M_N}{p_{N_z}}\right)}\right)\;\;\longrightarrow\;\; \half\,\frac{\Delta l}{\tau_N}\frac{M_N}{p_{N_z}}\,,
\end{eqnarray}
the limit is valid for our interesting case of small mixing, when the HNL decay length exceeds the length of the decay volume (and most of HNLs safely leave the detector volume); the most relevant quantity here is the HNL life-time $\tau_N$.  
To obtain the number $N_{decay}$ of  HNLs which decay inside the decay volume we count all of the HNL which reach the detector $N_{detector}$ with a weight $P$ from eq. \eqref{eq:P}. 
We present the ratios of $N_{decay}/N_{detector}$ in Fig. \ref{fig:N_decay_non} (left panel), 
setting $|U_{\alpha}|^2=10^{-8}$ as a reference number in the HNL lifetime $\tau_N$, which is calculated as explained below.    

\begin{figure}[!htb]
\centerline{
\includegraphics[width=0.45\textwidth]{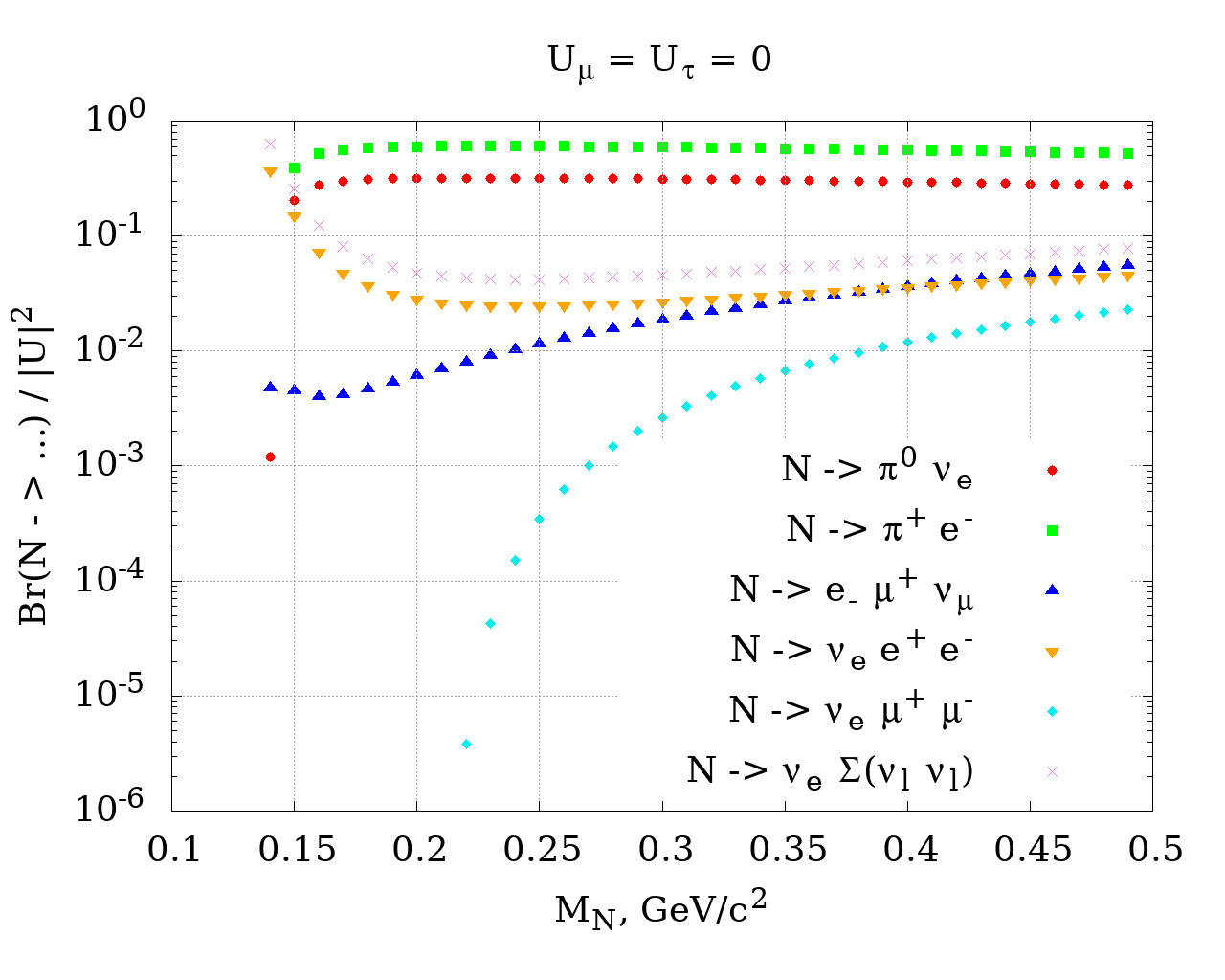}
\hskip 0.05\textwidth
\includegraphics[width=0.45\textwidth]{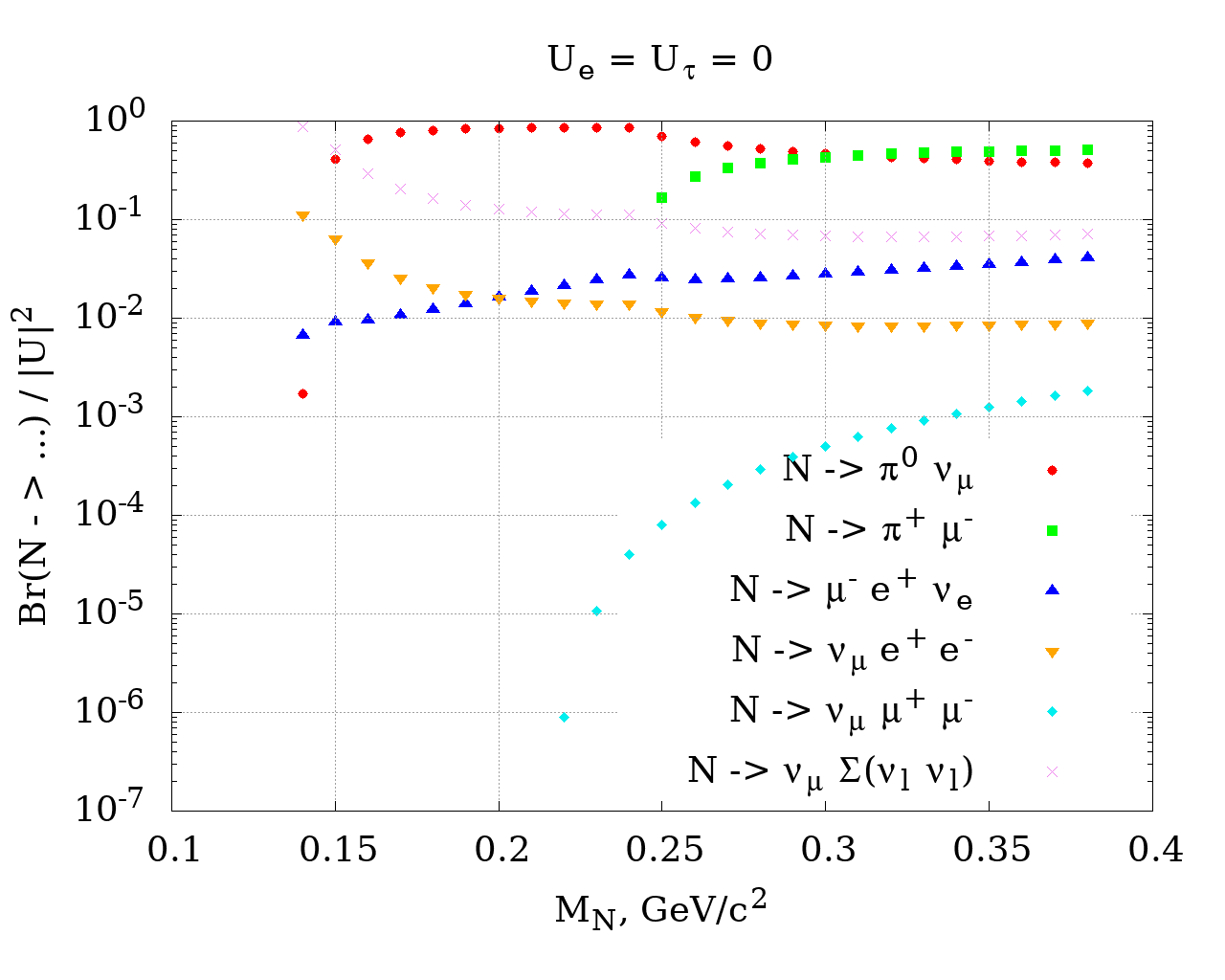}
}
\caption{Sterile neutrino decay modes in case of pure electronic (left panel) and pure muonic (right panel) mixing. In this case the branching ratios do not depend on the mixing parameter $|U_{\alpha}|^2$.}
\label{fig:Br}
\end{figure}

{\bf 7.} HNLs mix with active neutrinos, and hence decay into the SM particles due to weak interactions. The formulas for HNL decay rates can be found in Ref.\,\cite{Gorbunov:2007ak}. The branching ratios relevant for the interesting mass range  are presented in Fig.\,\ref{fig:Br}.  
One observes that in HNL decays as well as in their production, the two-body decay modes, when kinematically open, dominate. Moreover, the 2-body decay mode into charged particles dominates over that into neutral particles (except for the threshold regions). Thus, for the signal events we consider only these charged modes ($\pi^\pm l^\mp$, where $l$ is electron and muon), when kinematically open, also because they are treated presently at SHiP as the background-free ones. When 2-body decays are kinematically forbidden, we consider pure leptonic modes with two charged leptons, $N\to l^+_\alpha l^-_\beta \nu$. They are also treated as background free at SHiP \cite{SHiP:2018xqw}.   

The HNL lifetime for the reference mixing $|U_{\alpha 4}|^2=10^{-8}$ is shown on the right panel of Fig.\,\ref{fig:N_decay_non}. One finds that the HNL decay length decreases with its mass, but for the reference level of mixing it always vastly exceeds length of the SHiP decay volume, which justifies the limit \eqref{eq:P}. 

\begin{figure}[!htb]
\begin{center}
\includegraphics[width=0.45\textwidth]{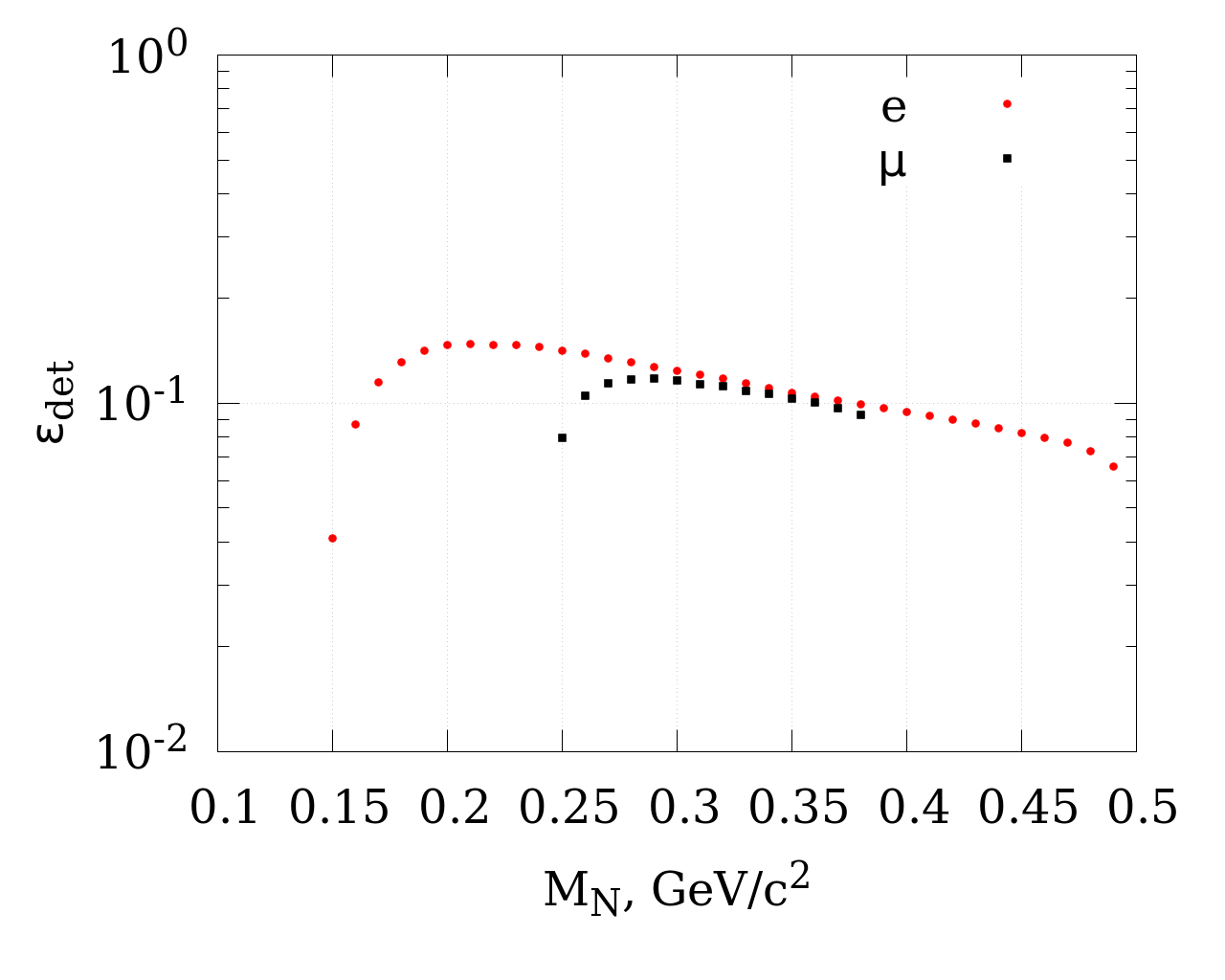}
\hskip 0.05\textwidth
\includegraphics[width=0.45\textwidth]{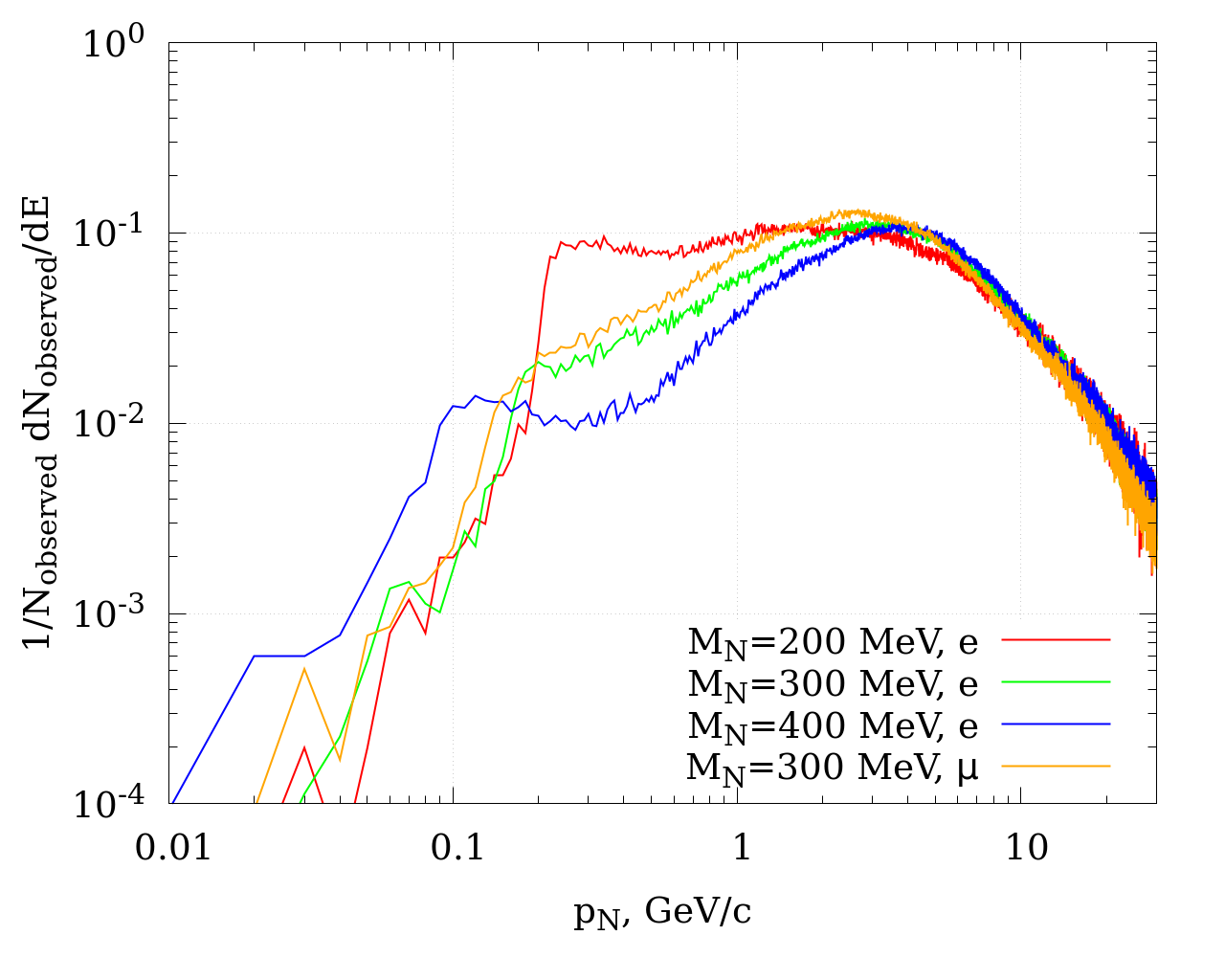}
\caption{Dependence of the geometrical acceptance of the detector for the HNLs produced from charged kaons on the sterile neutrino mass (left panel).
The spectrum of the produced by charged kaons sterile neutrinos, which decay products pass the geometrical criteria (right panel). 
}
\label{fig:N_detectable_non}
\end{center}
\end{figure}

\begin{figure}[!htb]
\begin{center}
\includegraphics[width=0.45\textwidth]{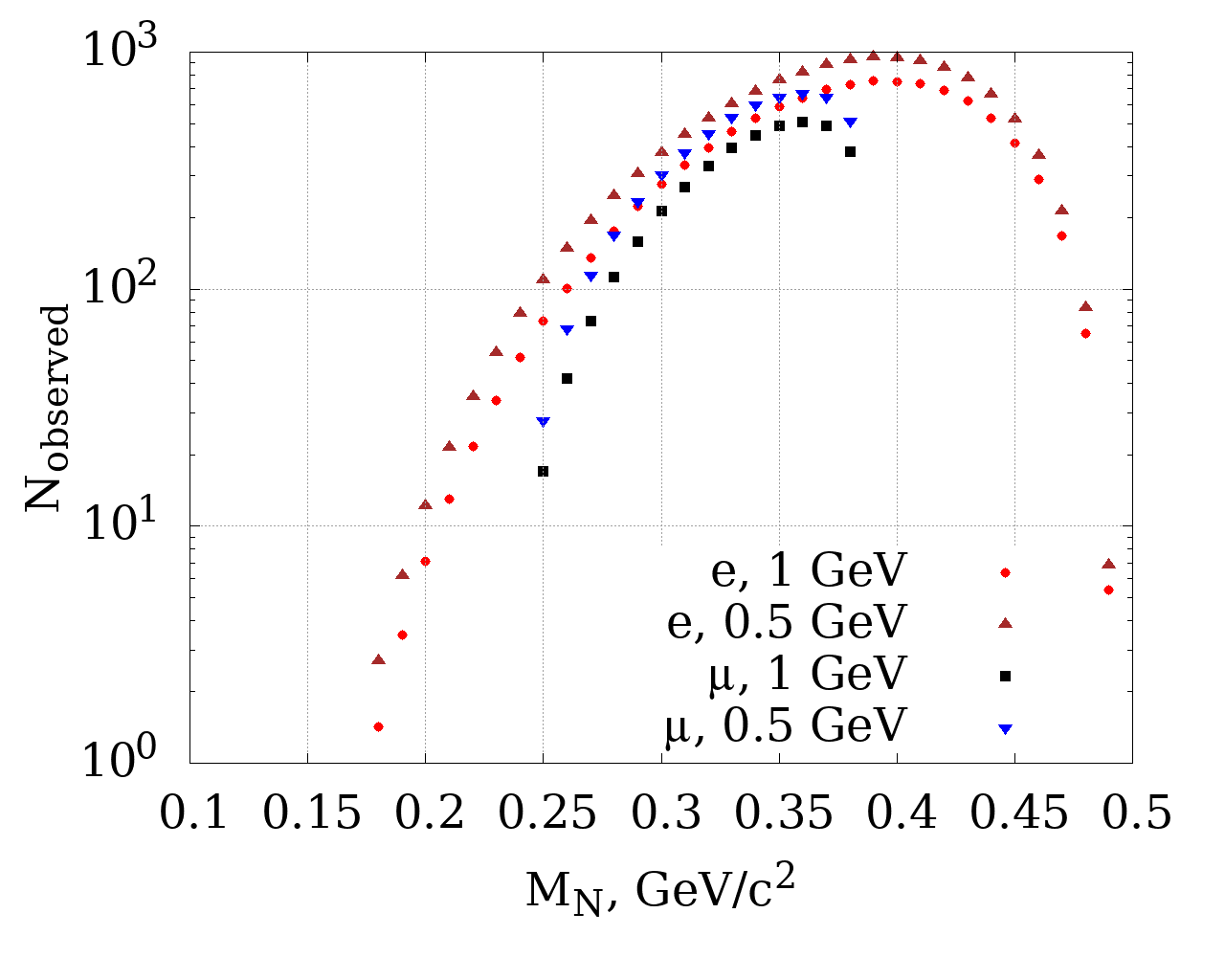}
\hskip 0.05\textwidth
\includegraphics[width=0.45\textwidth]{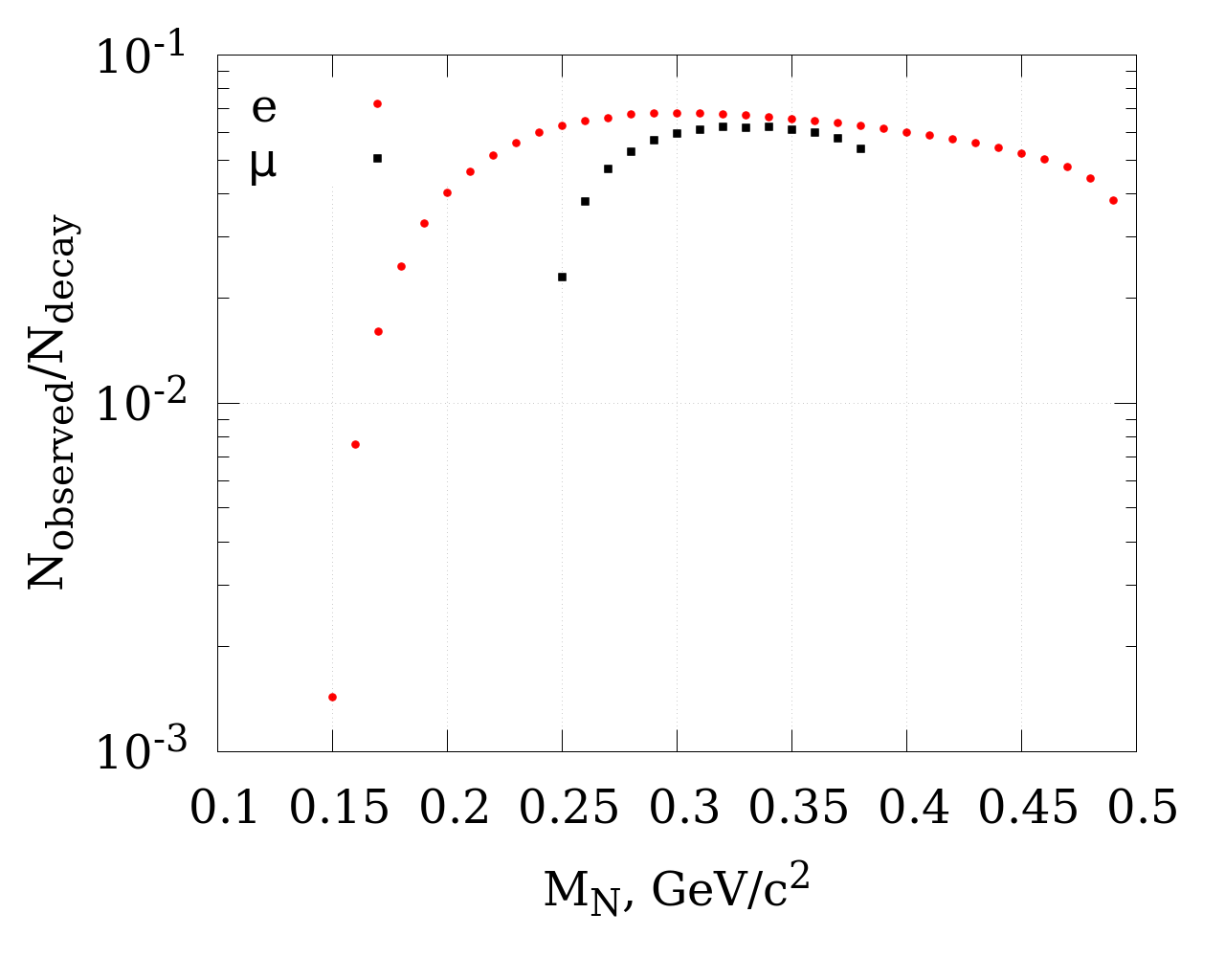}
\caption{Dependence of the  number of signal events from HNL on $M_N$ for two chosen cuts on the 3-momenta of HNL decay products (left panel). 
Dependence of the fraction of the observed sterile neutrinos produced from charged kaons on the sterile neutrino mass for the 1 GeV/c threshold of  the 3-momenta of HNL decay products (right panel).
}
\label{fig:1GeV}
\end{center}
\end{figure}

To be observed, the charged particles must cross the far end of the decay volume to enter the decay spectrometer installed there that makes it possible   to  identify  the particle type and to measure its energy. Moreover, its presently accepted operation scheme is supposed to be almost 100\% effective, if at least two charged particles from the decay have 3-momenta exceeding 1\,GeV/c.  
These constraints on the 3-momenta of the HNL decay products we further implement in our study applying the procedure above we use for the HNL trajectories. We briefly describe it here for the 2-body decay modes. 
Namely, we replace $N$ with $\pi$ and $K$ with $N$ in \eqref{eq:2body} and \eqref{p_Nx}, \eqref{p_Nz}.
In \eqref{eq:xN_yN} at the moment $z_\pi = L + \Delta l$ we replace starting point $\{x_1,y_1,z_1\}$ with $\{x_N,y_N,z_N\}$ from \eqref{eq:xN} - \eqref{eq:zN}. In this way we obtain the pion coordinates at the moment it reaches the  end of the decay volume $\{x_\pi,y_\pi,z_\pi\}$.
Taking into account that $\vec{p}_N = \vec{p}_\pi + \vec{p}_l$, similarly we obtain $\{x_l,y_l,z_l\}$, $z_l= L + \Delta l$. Both pion and charged lepton cross the backside of the decay volume  if $|x_\pi|< \Delta w_2, |x_l|< \Delta w_2$ and $|y_\pi| < \Delta h_2, |y_l| < \Delta h_2$.
In the same way as we do above with $N_{detector}$, we count all sterile neutrinos that pass these {\it geometrical criteria} and call the resulting number $N_{observed}$. 
Thus we find the geometrical acceptance of the detector $\epsilon_{det} \equiv N_{observed}/N_{decay}$ for the HNLs decaying in the decay volume. 
We present the geometrical acceptance of the decaying sterile neutrinos on the  left panel of Fig.\,\ref{fig:N_detectable_non}.
They only mildly depend on the HNL mass.  
The spectra of those HNLs, which decay products fulfill the geometrical criteria in the SHiP experiments are given on the right panel of  Fig.\,\ref{fig:N_detectable_non}. Finally, we apply the cut on the 3-momenta of the charged particles the HNL decay to, $\pi^\pm$ and $l^\mp$, obeying the geometrical criteria. This is rather constraining criterion, that we illustrate on the left panel of Fig.\,\ref{fig:1GeV} 
for the two chosen values of 0.5\,GeV/c and 1\,GeV/c. One observes that a reduction of the detection threshold would  noticeably increase the number of signal events. The presently accepted by the SHiP collaboration momentum threshold is 1\,GeV/c. The results for the HNL fraction passed both the geometrical and momentum constraints are shown on the right panel of Fig.\,\ref{fig:1GeV}.  Both populations of a light and a heavy HNL from the kinematically open mass range significantly suffer from the energy cut, and even HNL of mass in the central part, $M_N\sim 300$\,MeV, lose one order of magnitude in the signal statistics.

We arrive at the similar conclusions for the leptonic decay modes.

\begin{figure}[!htb]
\begin{center}
\includegraphics[width=0.7\textwidth]{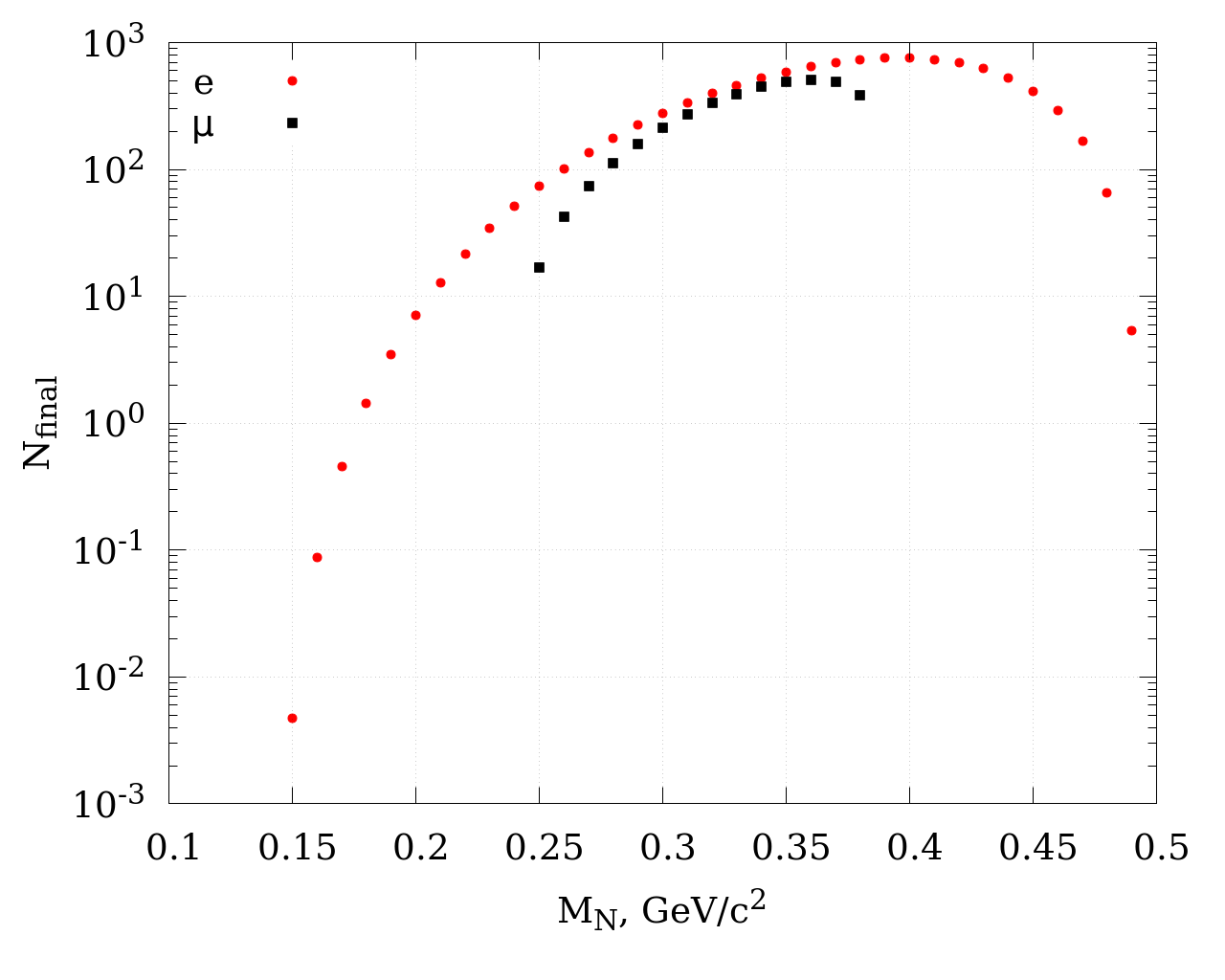}
\caption{The total number of the observed sterile neutrinos (2-body decay modes only) produced by charged kaons as a function of  the sterile neutrino mass; SHiP statistics is $N_{POT}=2\times 10^{20}$, see Tab.\,\ref{tab:initial}.}
\label{fig:N_final_non}
\end{center}
\end{figure}

{\bf 8.} 
To obtain the number $N_{final}$ of the observed sterile neutrinos that decay inside the decay volume we count all of the observed sterile neutrinos $N_{observed}$ with a weight $P$ from eq.\,\eqref{eq:P}.
We present the final number of the observed sterile neutrinos $N_{final}$ in Fig.\,\ref{fig:N_final_non} for 2-body decay channels.

\begin{figure}[!htb]
\begin{center}
\includegraphics[width=0.7\textwidth]{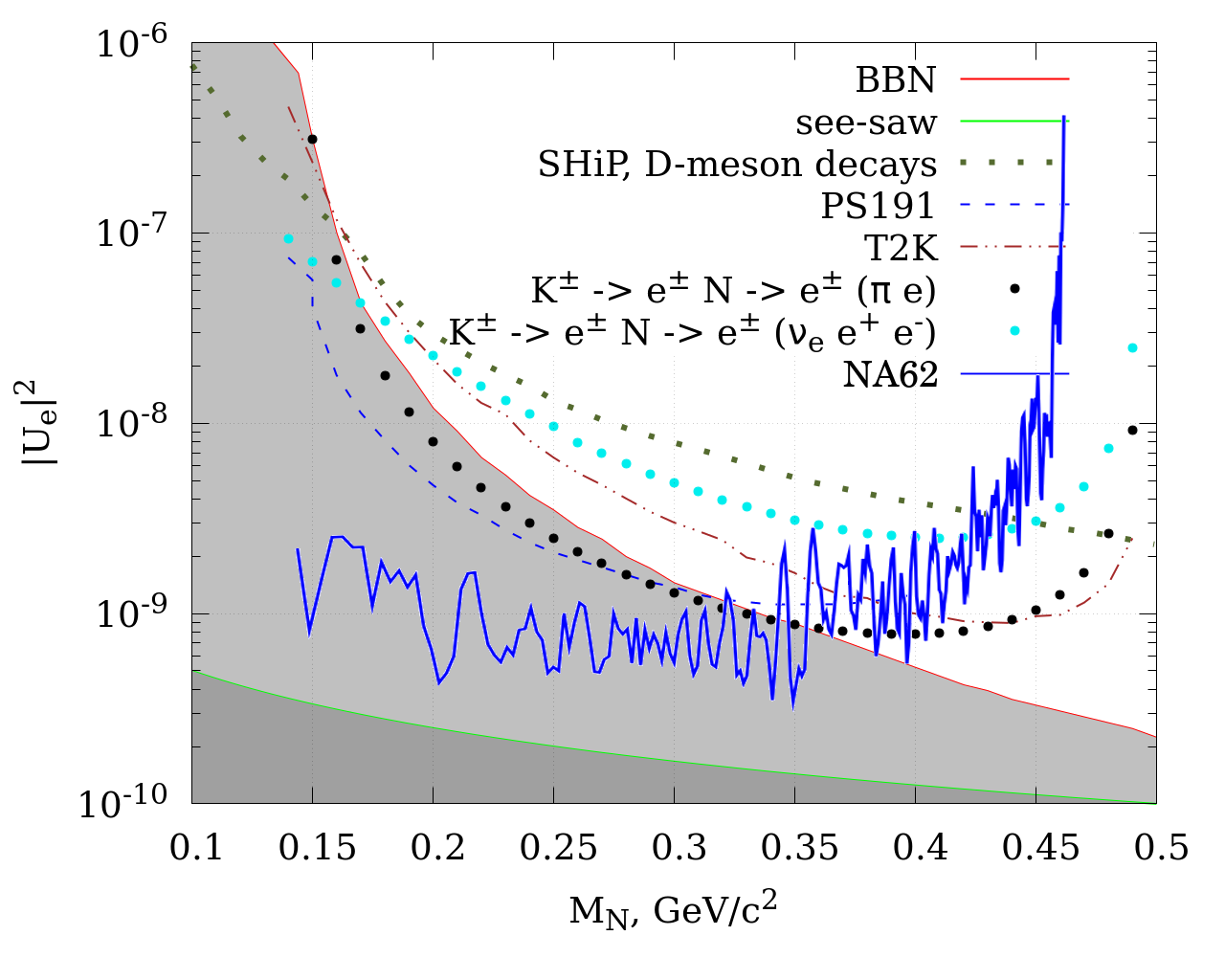}
\caption{Present limits on the HNL mixing with electron neutrino and expected sensitivity of the SHiP experiment with statistics $N_{POT}=2\times 10^{20}$, see Tab.\,\ref{tab:initial}. Also shown are the results obtained in PS191~\cite{Bernardi:1987ek},  and T2K~\cite{Abe:2019kgx} (within Feldman--Cousins approach) and recent results of NA62\,\cite{NA62:2020mcv}.}
\label{fig:sensitivity_e}
\end{center}
\end{figure}

\begin{figure}[!htb]
\begin{center}
\includegraphics[width=0.7\textwidth]{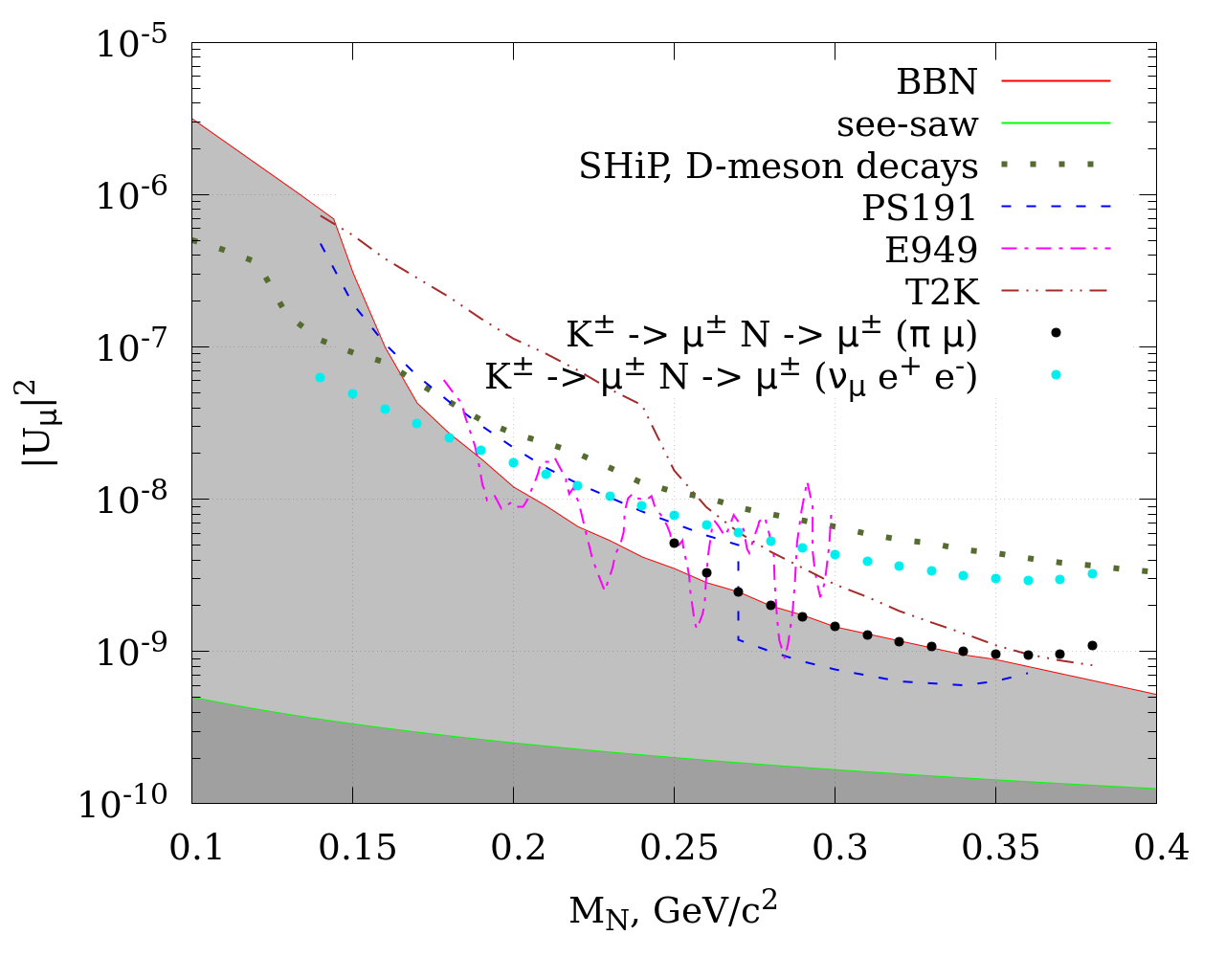}
\caption{Present limits on the HNL mixing with muon neutrino and expected sensitivity of the SHiP experiment with statistics $N_{POT}=2\times 10^{20}$, see Tab.\,\ref{tab:initial}. Also shown are the results obtained in PS191~\cite{Bernardi:1987ek}, E949~\cite{Artamonov:2014urb}, and T2K~\cite{Abe:2019kgx}.}
\label{fig:sensitivity_mu}
\end{center}
\end{figure}

\begin{figure}[!htb]
\begin{center}
\includegraphics[width=0.45\textwidth]{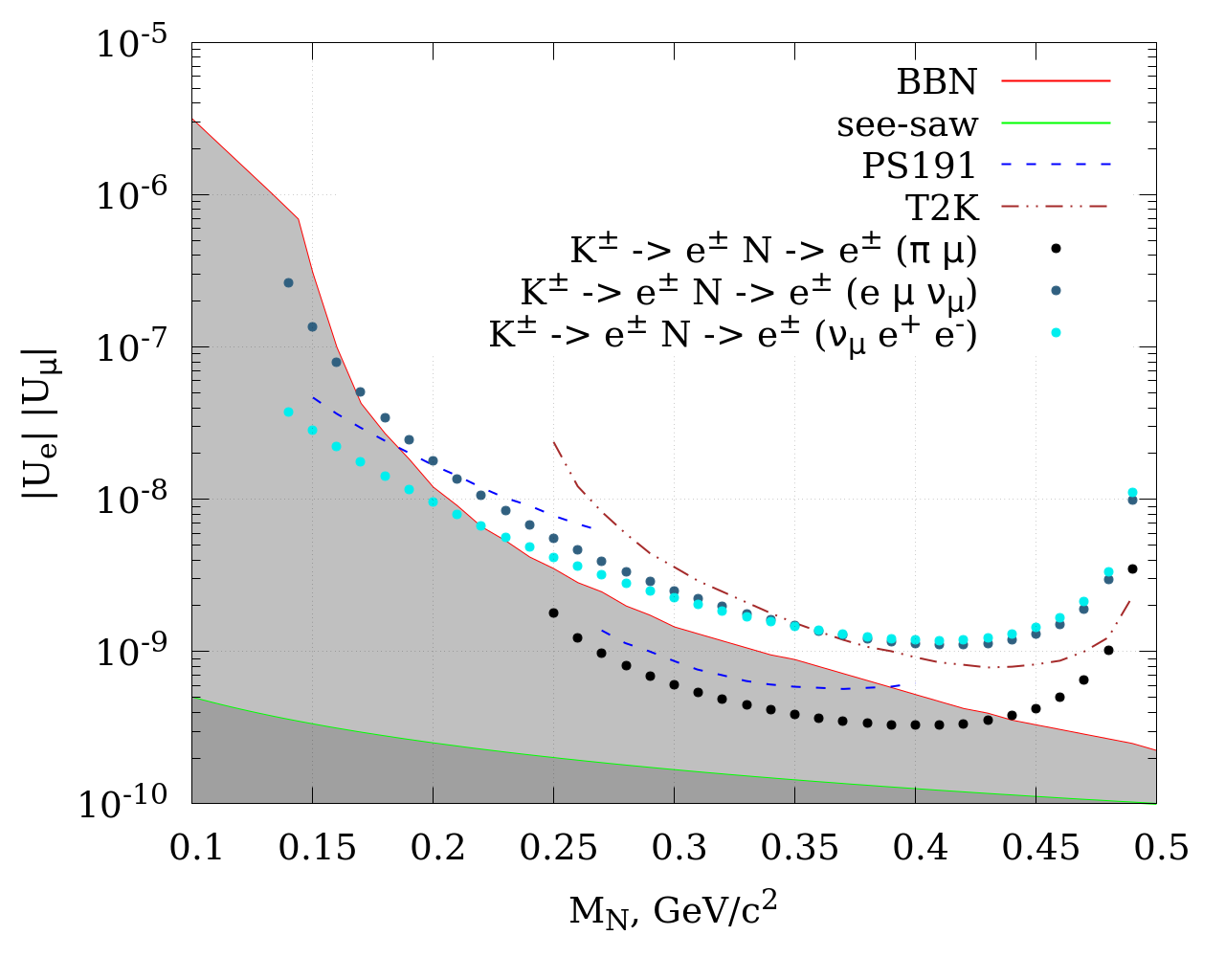}
\hskip 0.05\textwidth 
\includegraphics[width=0.45\textwidth]{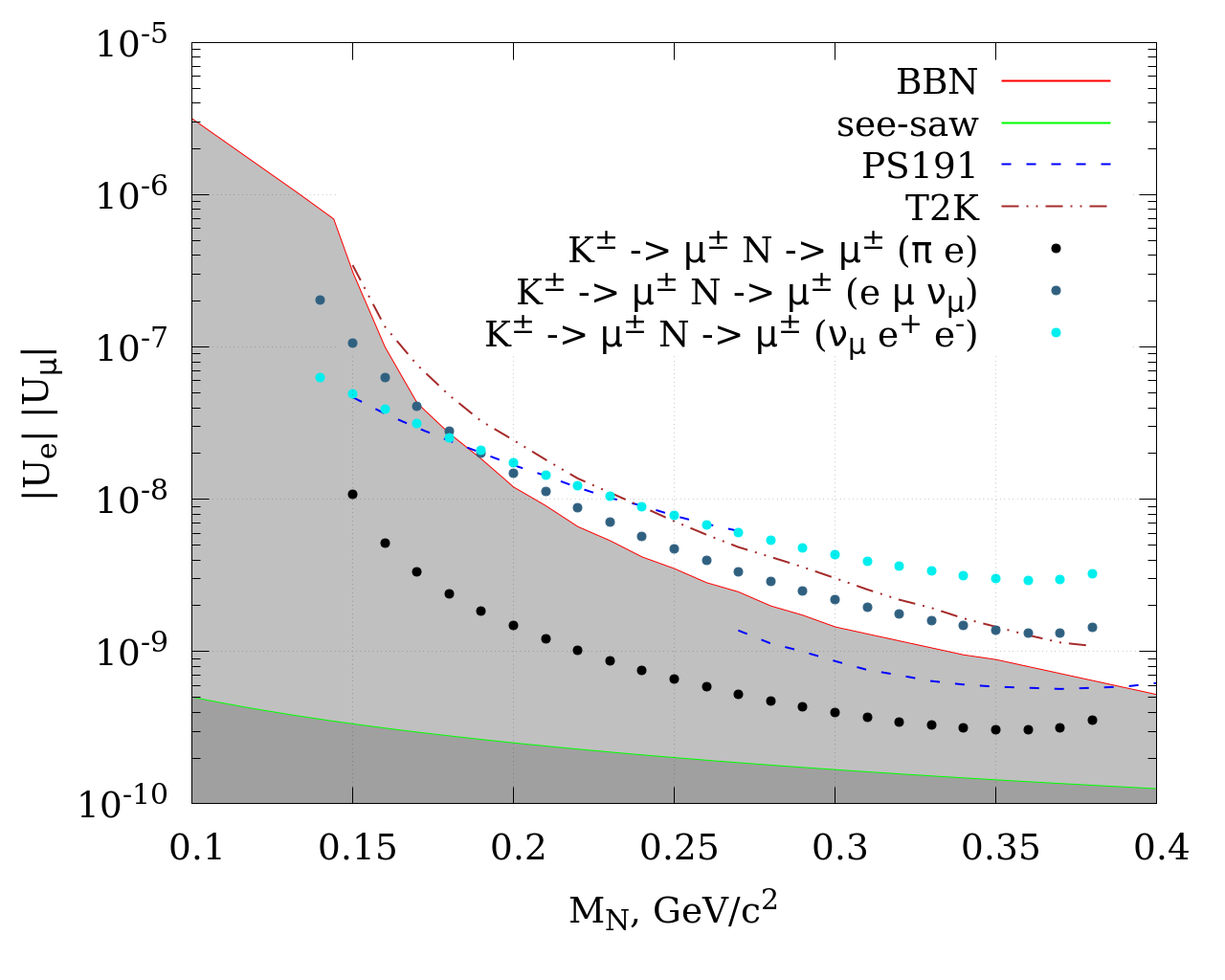}
\caption{Present limits on the combination $|U_e U_\mu|$ (left panel) and $|U_\mu U_e|$ (right panel) of HNL mixing with muon and electron neutrinos and expected sensitivity of the SHiP experiment with statistics $N_{POT}=2\times 10^{20}$, see Tab.\,\ref{tab:initial}. The parameter products are the same, but the corresponding observables are different, so we present them separately for convenience.}
\label{fig:sensitivity_emu}
\end{center}
\end{figure}

\begin{figure}[!htb]
\begin{center}
\includegraphics[width=0.45\textwidth]{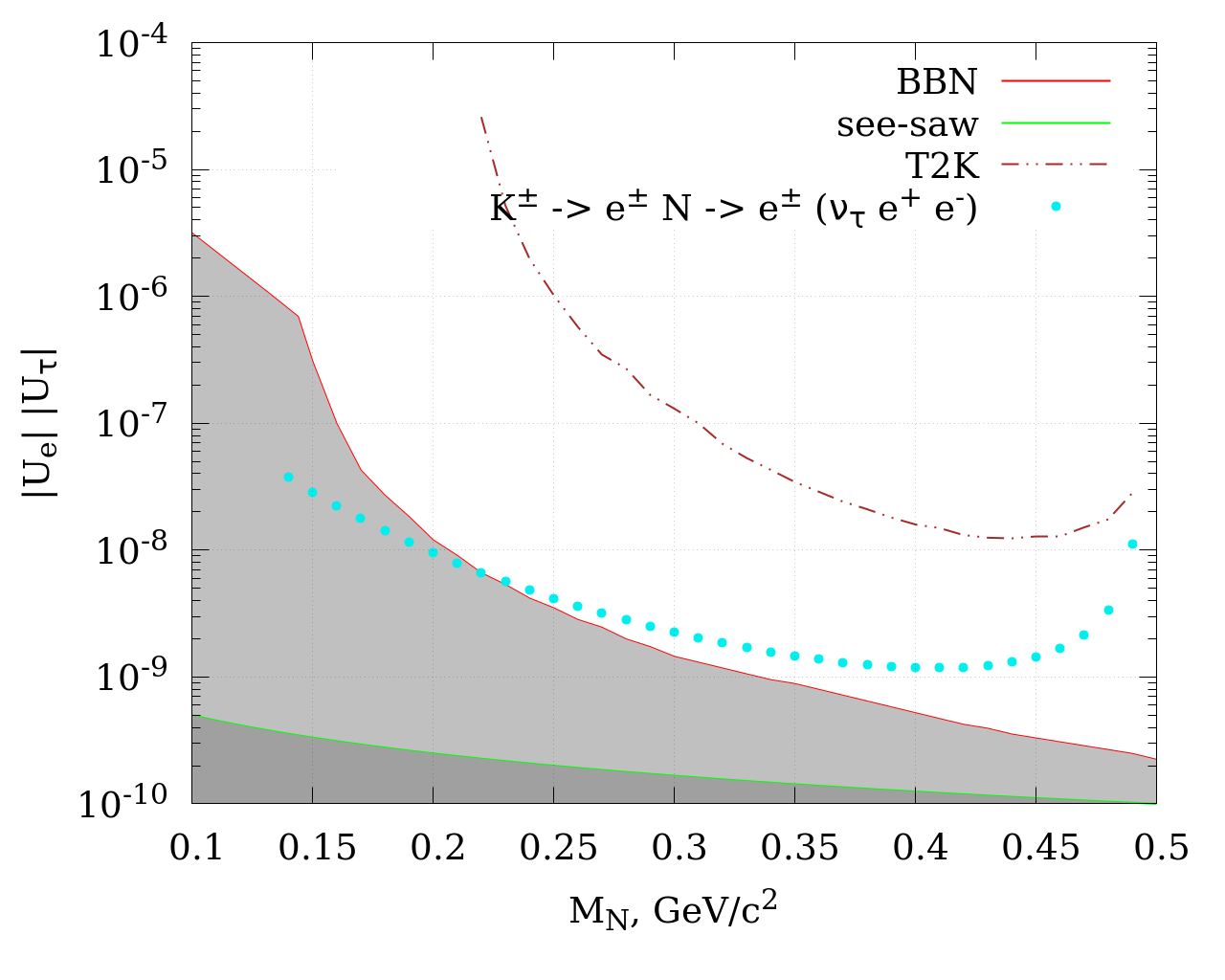}
\hskip 0.05\textwidth 
\includegraphics[width=0.45\textwidth]{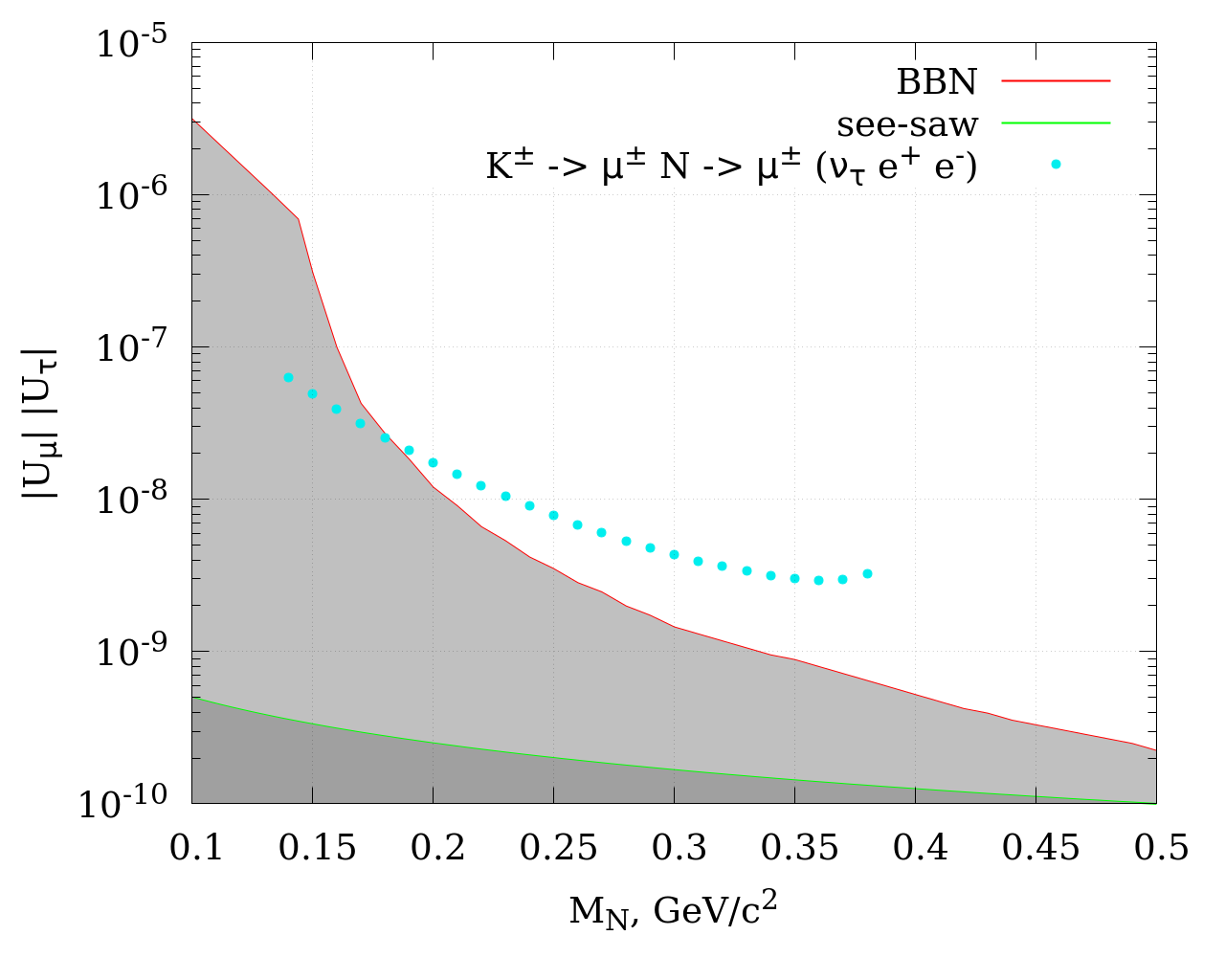}
\caption{Present limits on the combination $|U_e U_\tau|$ (left panel, assuming $|U_e|\ll |U_\tau|$) and $|U_\mu U_\tau|$ (right panel, assuming $|U_\mu|\ll |U_\tau|$) of HNL and expected sensitivity of the SHiP experiment with statistics $N_{POT}=2\times 10^{20}$, see Tab.\,\ref{tab:initial}.}
\label{fig:sensitivity_tau}
\end{center}
\end{figure}

Accepting the background-free scheme at SHiP, we ask $N_{final}\leq2.3$ to put the 90\% CL upper limit on the active-sterile mixing at a given HNL mass, assuming no evidence \emph{for the chosen HNL signature.}  Recall, that we accepted no suppression, i.e. $|U_\alpha|^2=1$, for the productions and use $|U_\alpha|^2 = 10^{-8}$ for HNL decays as the reference number to make numerical estimates and present the results on plots. Since the number of events depends on the product of our two reference numbers, the correct limit in our scheme is given as 
\begin{equation}
|U_\alpha U_\beta|^2 = \sqrt{\frac{2.3}{N_{final}}}\times 10^{-8}\,,
\end{equation} 
where $\alpha$ and $\beta$ refer to the mixings responsible for HNL production and decay, respectively. Moreover, the limits are placed investigating a particular signature -- decay mode -- and we consider them separately. 
The corresponding SHiP sensitivity is presented in Figs.\ref{fig:sensitivity_e}, \ref{fig:sensitivity_mu}, \ref{fig:sensitivity_emu} and \ref{fig:sensitivity_tau}  
together with SHiP sensitivity based on D-meson decays\,\cite{SHiP:2018xqw} and  previous limits on the sterile-active 
mixing from direct searches, from cosmology (grey regions: Big Bang  Nucleosynthesis constraints are scanned from Fig.\,4.10 of Ref.\,\cite{Alekhin:2015byh} which in the interesting mass range is repeated from the restriction of $\tau_s<0.1$\,sec in Fig. 1 of Ref.\,\cite{Boyarsky:2009ix}) and the reference line for the see-saw mechanism relation, $U^2\approx\frac{\sqrt{m_{atm}^2}}{M_N} \approx 0.05\,\text{eV} / M_N$.  One observes that the SHiP experiment is capable of fully exploring the central parts of kinematically available regions above the cosmological BBN constraint (the regions shown in grey), if the HNL lifetime is saturated by the same mixing.  Remarkably, the see-saw line refers approximately to the \emph{minimal mixing} at which the corresponding sterile neutrino contributes sufficiently large amount to active neutrino masses. The region \emph{below the BBN line} is forbidden from cosmology. Therefore, if the direct searches reveal negative results, \emph{this will close the simple seesaw scheme for the light HNL} provided BBN limits are solid.  What will remain is the hierarchical pattern: the dominant mixing with tau-neutrino, $|U_\tau|^2\gg |U_\mu|^2,\,|U_e|^2$ may ensure HNL decays faster than 0.1\,s because of relatively larger $|U_\tau|^2$, which is not directly tested at SHiP.

Note, that in our study we treat HNL as Majorana particles, while they are considered to be Dirac in PS191 and in E949 experimental analyses.  
Therefore, to present these results on our plots we have divided their limits by square root of two. 
Moreover, the original analysis of PS191 collaboration is based only on  HNL interactions via charged currents: the neutral current contributions to all processes with HNL have been ignored. Following the analysis of Ref.\,\cite{Ruchayskiy:2011aa} we include the missed contributions from the neutral currents and implement all the relevant corrections for PS191 limits presented in 
Figs.\,\ref{fig:sensitivity_e}-\ref{fig:sensitivity_emu}.


{\bf 9.} 
To conclude, in this paper we calculate the signal events from hypothetical heavy neutral leptons emerged from decays of charged kaons  inside the Mo-Tn target of the SHiP experiment. We find this source of HNL is more promising for masses in the range of 150 - 500 MeV than previously considered decays of  D-mesons\,\cite{SHiP:2018xqw}, so that it allows SHiP to fully explore the large part of phenomenologically and cosmologically viable region of sterile-active mixing in models with HNLs lighter than kaons (except the case where mixing with tau neutrino strongly dominates).   

In this study we account for only the main production channels and HNL decay modes. It seems that corrections from the subdominant processes do not change much. First, consider decays. The dominant decay modes are 2-body decays, which rates are amplified by the chirality flip factor $\propto M^2_N$. The 3-body decays are suppressed by the   phase space factor and are kinematically forbidden for a heavy HNL. The low-threshold three-body leptonic mode, $N\to e^+e^-\nu$ becomes competitive only for a light HNL, $M_N\lesssim m_\pi$, see Fig.\,\ref{fig:Br}, but then pions must be introduced to the analysis as a possible source of such an HNL, that is beyond the scope of this paper. Second, turn to the production. There, the strange baryons weakly decaying to ordinary ones would also contribute \cite{Ramazanov:2008ph} but only to a light HNL of $M_N\lesssim 100$\,MeV, where again, the pions must be considered. Notice, neutral kaons may decay into a heavier HNL (e.g. $K^0\to \pi e N$), but with strongly suppressed rates, see Ref.\,\cite{Gorbunov:2007ak}.     

The signal rate of HNL with $m_\pi < M_N< M_K$ can be increased by a factor of about 2 with a  lower cut (say, 0.5\,GeV/c) on the 3-momenta of the HNL decay products, see the left panel of Fig.\,\ref{fig:1GeV}.  Further, in that case one can also investigate the decay events when one of the charged particles goes backwards (upstream, towards the target) and use the tracker system and the neutrino detector in front of the SHiP decay volume to identify the particle type and measure its momentum. An ultimate (but probably unrealistic) option is 
that sufficiently low cuts would allow SHiP to register the HNL from kaon decays at rest, which statistics exceeds that of decays in flight by an order of magnitude. 
The detector geometric acceptance, see left panel of Fig.\,\ref{fig:N_detectable_non}, can be possibly increased by using the veto system on the surface of the decay volume to help in detecting the HNL decay event: time and position of crossing by  an outgoing charged particle. 

The overall SHiP sensitivity to models with HNL can possibly be increased in this way, but modestly. The number of signal events scales as $|U_\alpha|^2\times |U_\beta|^2$, and these steps can 
help to pull down the expected limits by a factor 
of 2,    farther probing the region around the cosmological BBN constraint. The next major scientific goal of direct searches --  testing the region of the seesaw line $|U_\alpha|^2\sim m_\nu/M_N$, see  Figs.\,\ref{fig:sensitivity_mu}-\ref{fig:sensitivity_emu} -- requires an increase of  a factor of hundred  in the number of signal events and will be a great challenge for the  next generation experiments.     

\vskip 0.5cm

The work was supported in part by the RSF Grant No. 17-12-01547. 
\bibliographystyle{utphys}
\bibliography{HNL-from-kaons}
\end{document}